\pdfoutput=1

\documentclass[runningheads,a4paper]{llncs}

\usepackage{textcomp}
\usepackage{amsmath,amssymb,amsfonts}
\usepackage{graphicx}

\usepackage{times}
\usepackage{xcolor}
\usepackage{soul}
\usepackage[utf8]{inputenc}
\usepackage[small]{caption}
\usepackage{url}

\urldef{\mailsa}\path|oota.subba@students.iiit.ac.in|
\urldef{\mailsb}\path|{naresh.manwani, raju.bapi}@iiit.ac.in|

\begin{document}

\mainmatter  

\title{fMRI Semantic Category Decoding using Linguistic
Encoding of Word Embeddings}


%
%
\author{Subba Reddy Oota$^{1}$%
\and Naresh Manwani$^{1}$\and Raju S. Bapi$^{1,2}$}
\authorrunning{Subba Reddy Oota, Naresh Manwani, Bapi Raju S.}

\institute{$^{1}$ International Institute of Information Technology , Hyderabad\\
$^{2}$ School of Computer and Information Sciences, University of Hyderabad, Hyderabad \\
India\\
\mailsa\\
\mailsb\\
}

%
%

\toctitle{Lecture Notes in Computer Science}
\tocauthor{Authors' Instructions}
\maketitle

\begin{abstract}
The dispute of how the human brain represents conceptual knowledge has been argued in many scientific fields. Brain imaging studies have shown that the spatial patterns of neural activation in the brain are correlated with thinking about different semantic categories of words (for example, tools, animals, and buildings) or when viewing the related pictures. In this paper, we present a computational model that learns to predict the neural activation captured in functional magnetic resonance imaging (fMRI) data of test words. Unlike the models with hand-crafted features that have been used in the literature, in this paper we propose a novel approach wherein decoding models are built with features extracted from popular linguistic encodings of Word2Vec, GloVe, Meta-Embeddings in conjunction with the empirical fMRI data associated with viewing several dozen concrete nouns. We compared these models with several other models that use word features extracted from FastText, Randomly-generated features, Mitchell's 25 features~\cite{mitchell:predicting}. The experimental results show that the predicted fMRI images using Meta-Embeddings meet the state-of-the-art performance. Although models with features from GloVe and Word2Vec predict fMRI images similar to the state-of-the-art model, model with features from Meta-Embeddings predicts significantly better. The proposed scheme that uses popular linguistic encoding offers a simple and easy approach for semantic decoding from fMRI experiments.
\end{abstract}

\section{Introduction}

How a human brain represents and organizes conceptual knowledge has been an open research problem that attracted researchers from various fields~\cite{Alfonso-Bradford:conceptual,mahon-alfonso:drives,DHA:neural,tong-michael:decoding}. In recent studies, the topic of exploring semantic representation in the human brain has attracted the attention of researchers from both neuroscience and computational linguistic fields. Using brain imaging studies Neuroscientists have shown that distinct spatial/temporal patterns of fMRI activity are associated with different stimuli such as face or scrambled face~\cite{VMJ:fmri}, semantic categories of pictures, including tools, animals, and buildings, playing a movie, etc.~\cite{howell:statistical,haxby:distributed,ishai:distributed,kanwisher:fusiform,carlson:patterns,cox:functional,hhg:distributed,polyn:category}. These experimental results postulate how the brain encodes meaning of words and knowledge of objects, including theories that meanings are encoded in the sensory-motor cortical areas~\cite{caramjenni:domain,crutch:spatial,samson:orthographic}. Such findings would also facilitate making predictions about breakdown in the function and their spatial location in different neurological disorders. Theoretical and empirical studies have been conducted to explore categorization of animate and inanimate objects and the brain representation of these semantic differences~\cite{cree:analyzing,mahon:orchestration}. Linguists have identified different semantic meanings corresponding to individual verbs as well as the types of a noun that can fill those semantic meanings, for example, WordNet~\cite{miller:introduction}, VerbNet~\cite{schuler:verbnet}, and BabelNet~\cite{navigli:babelnet}. Tom Mitchell's group at {CMU} pioneered studies that demonstrated common semantic representation for various nouns in terms of shared brain activation patterns across subjects~\cite{mitchell:predicting}. In~\cite{singh2007detection}, presented the idea of detecting the cognitive
state of a human subject based on the fMRI data by exploring different classification techniques.

\begin{figure*}[!htb]
\minipage{0.333\textwidth}
  \includegraphics[width=\linewidth]{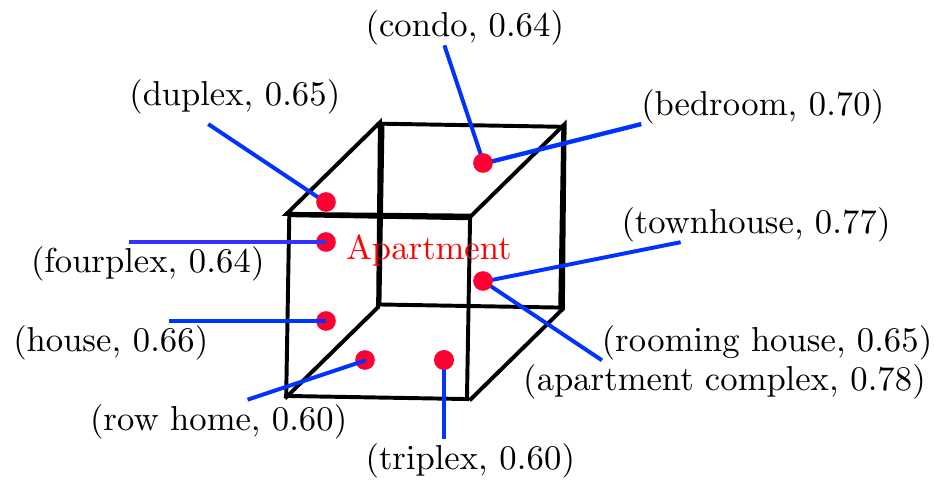}
  \vspace{-0.5cm}
\begin{center}

\end{center}
\endminipage\hfill
\minipage{0.333\textwidth}
  \includegraphics[width=\linewidth]{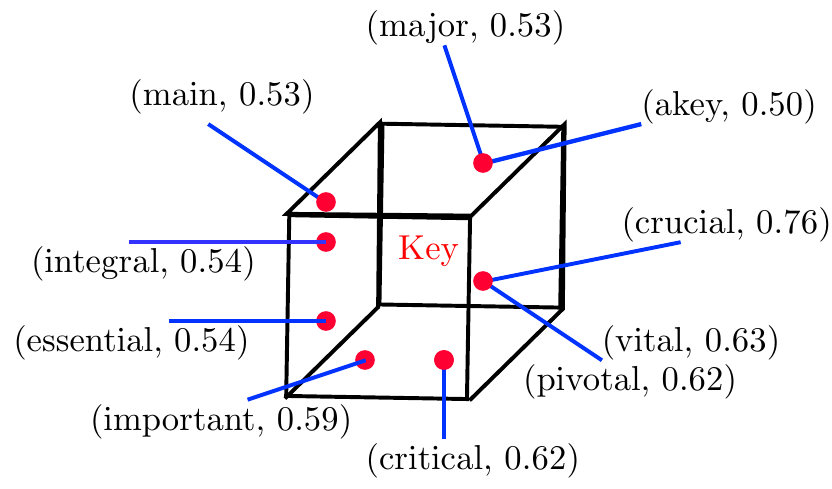}
  \vspace{-0.5cm}
\begin{center}
\textbf{(a) Word2Vec}
\end{center}
\endminipage\hfill
\minipage{0.333\textwidth}
  \includegraphics[width=\linewidth]{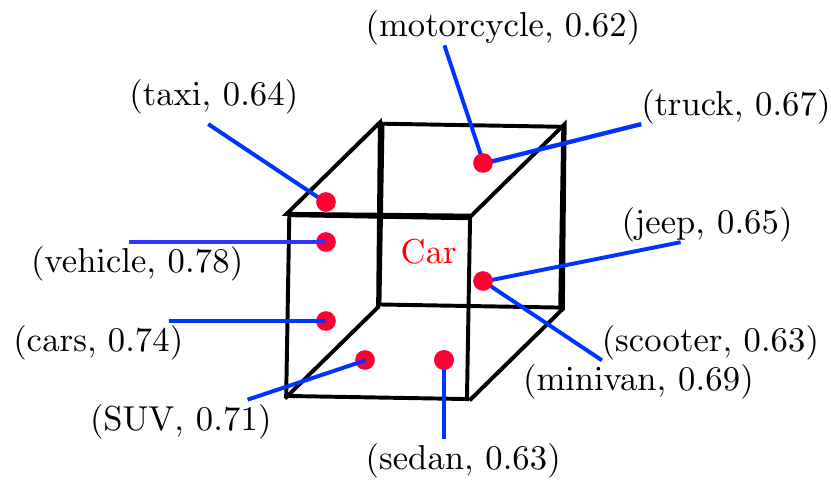}
 \vspace{-0.5cm}
\begin{center}

\end{center}
\endminipage
\vspace{0.1cm}
\minipage{0.333\textwidth}
  \includegraphics[width=\linewidth]{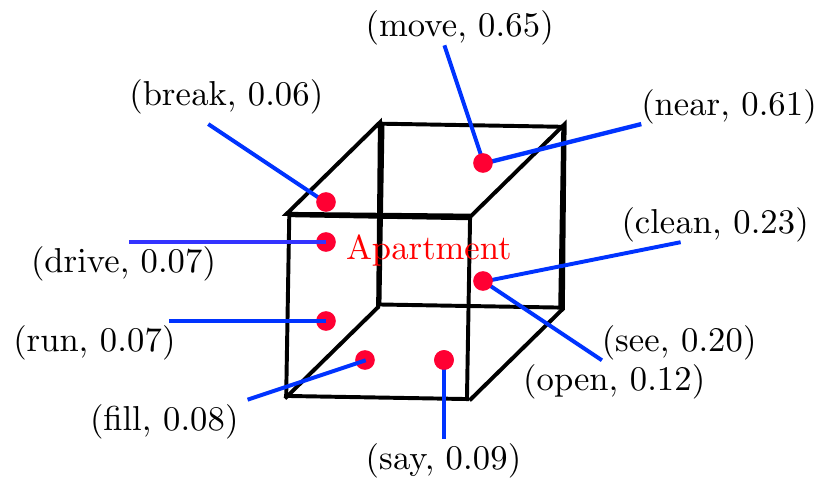}
  \vspace{-0.5cm}
\begin{center}

\end{center}
\endminipage\hfill
\minipage{0.333\textwidth}
  \includegraphics[width=\linewidth]{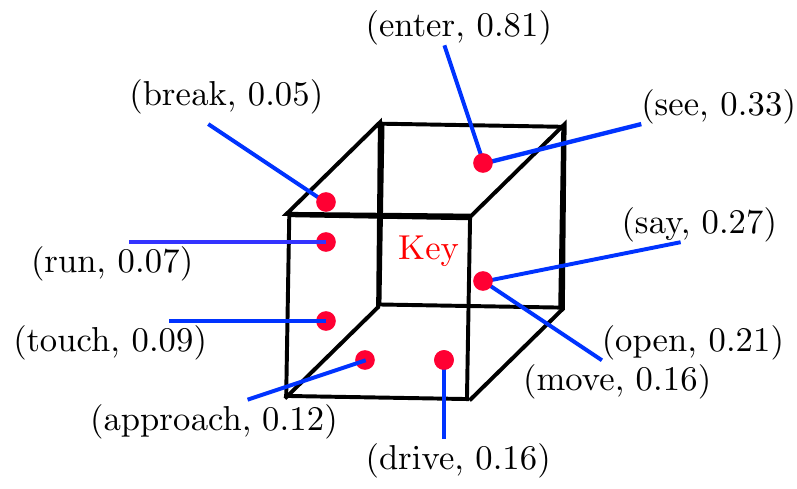}
  \vspace{-0.5cm}
\begin{center}
\textbf{(b) Mitchell's}
\end{center}
\endminipage\hfill
\minipage{0.333\textwidth}
  \includegraphics[width=\linewidth]{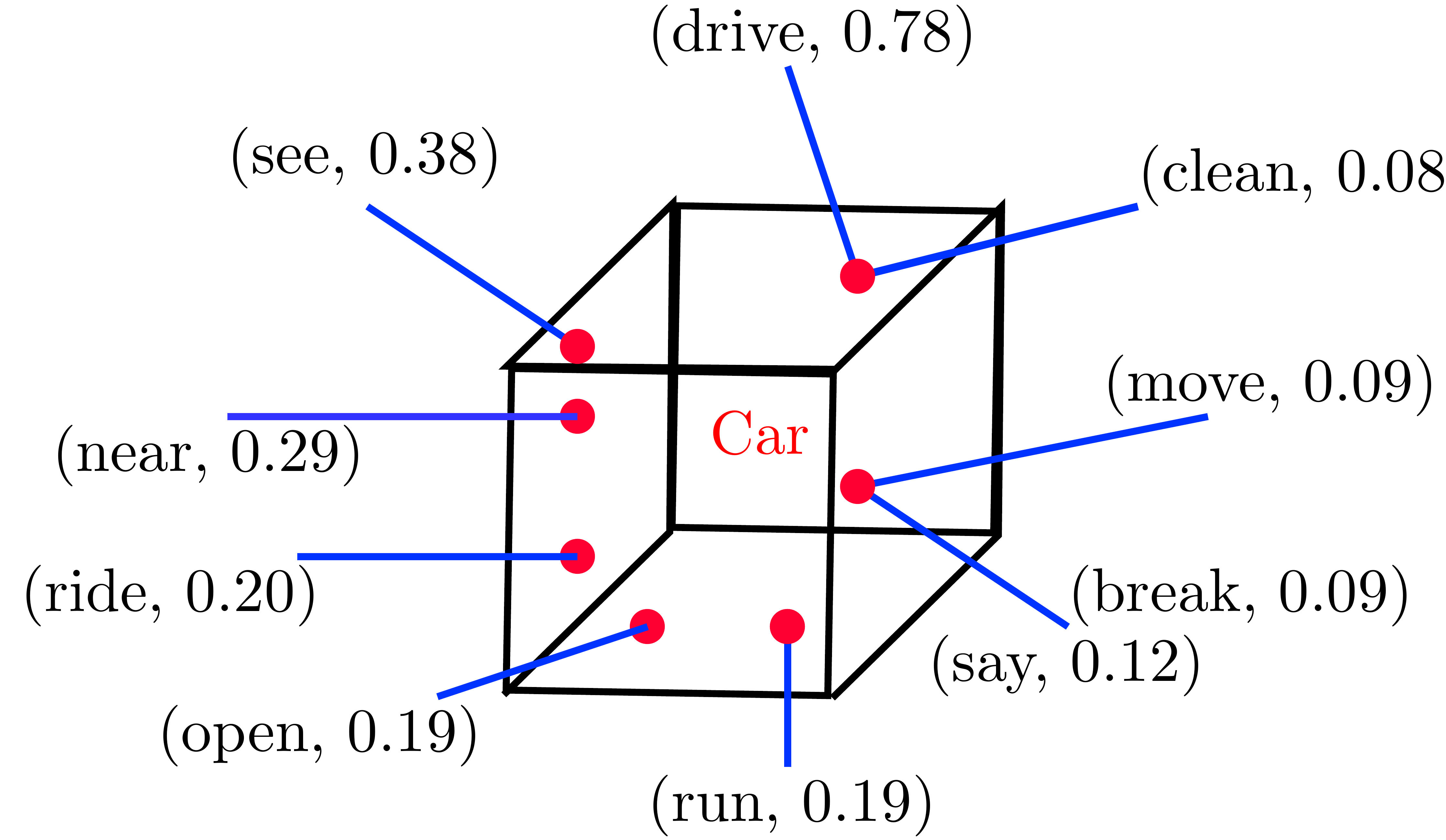}
 \vspace{-0.5cm}
\begin{center}

\end{center}
\endminipage
\vspace{0.1cm}
\minipage{0.333\textwidth}%
  \includegraphics[width=\linewidth]{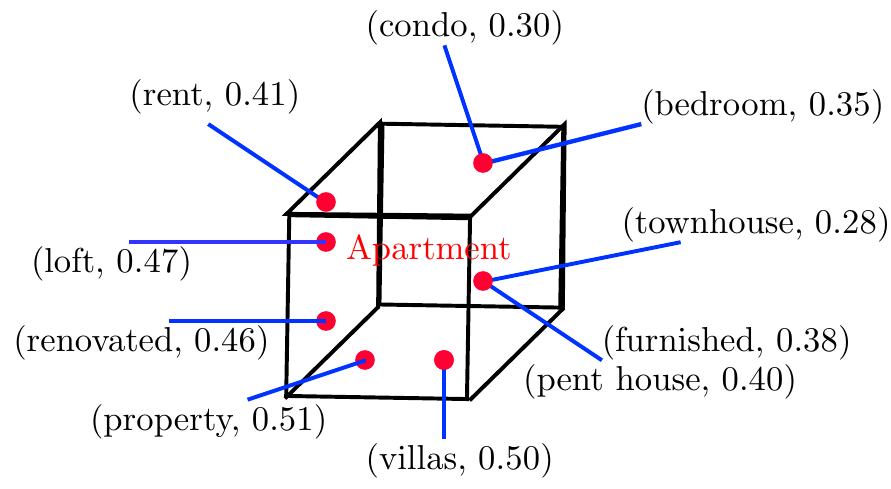}
  \vspace{-0.5cm}
\begin{center}

\end{center}
\endminipage\hfill 
\minipage{0.333\textwidth}%
  \includegraphics[width=\linewidth]{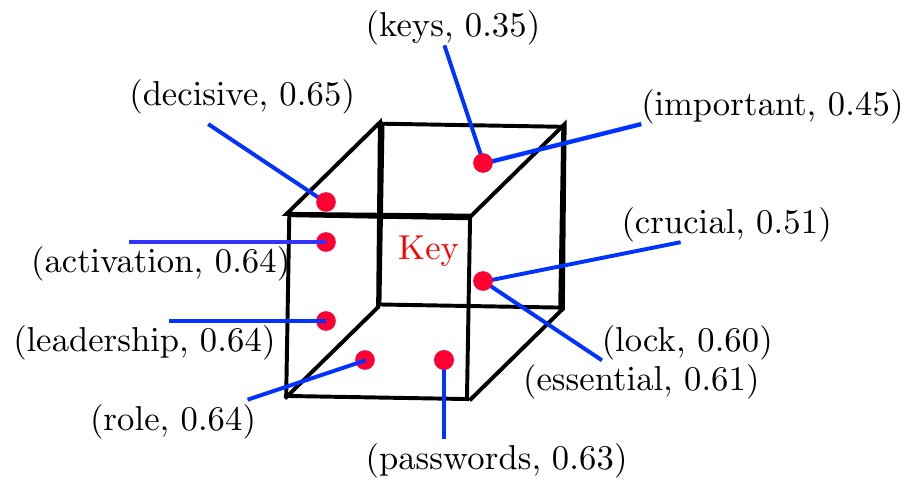}
  \vspace{-0.5cm}
\begin{center}
\textbf{(c) Meta-Embeddings}
\end{center}
\endminipage\hfill 
\minipage{0.333\textwidth}
  \includegraphics[width=\linewidth]{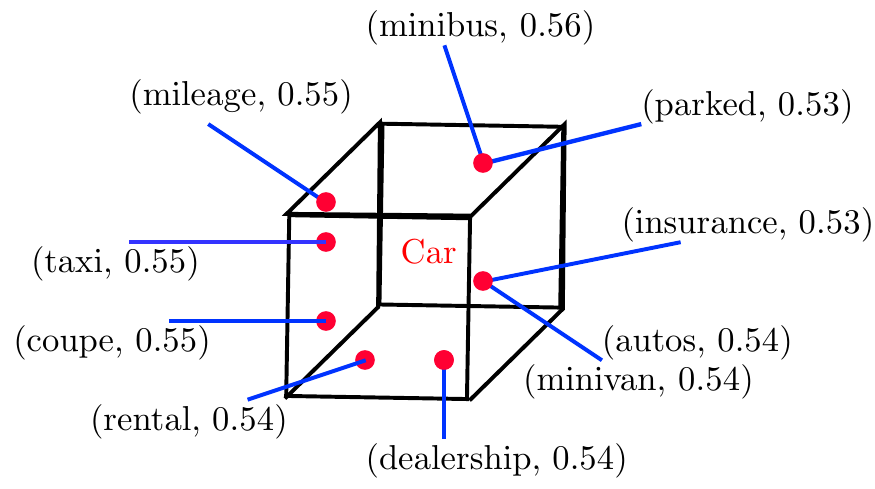}
  \vspace{-0.5cm}
\begin{center}

\end{center}
\endminipage
\vspace{0.1cm}
\minipage{0.333\textwidth}
  \includegraphics[width=\linewidth]{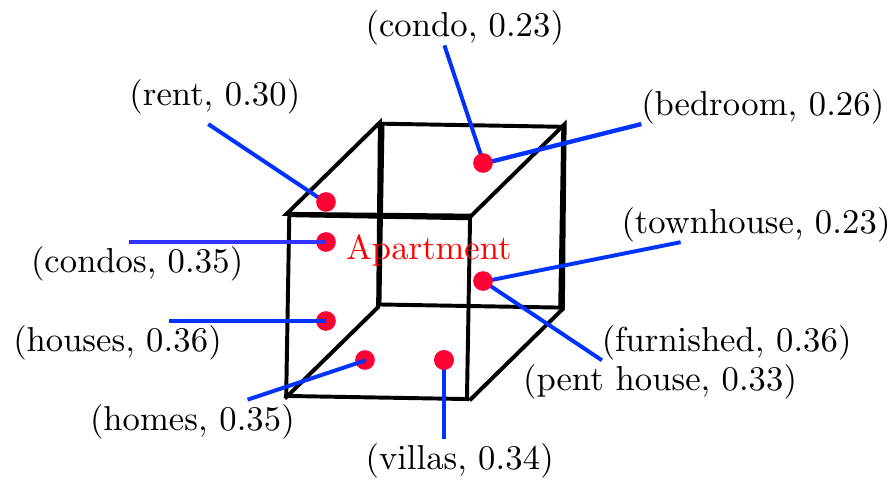}
 \vspace{-0.5cm}
\begin{center}

\end{center}
\endminipage\hfill
\minipage{0.333\textwidth}
  \includegraphics[width=\linewidth]{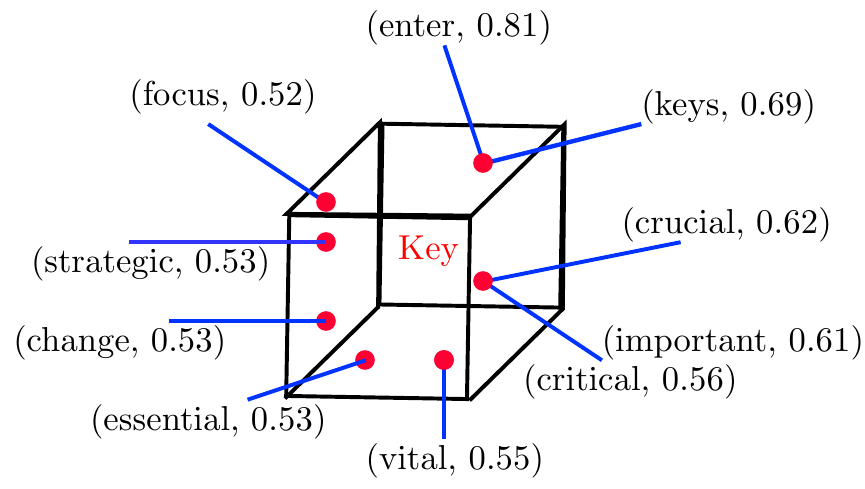}
 \vspace{-0.5cm}
\begin{center}
\textbf{(g) GloVe}
\end{center}
\endminipage\hfill
\minipage{0.333\textwidth}
  \includegraphics[width=\linewidth]{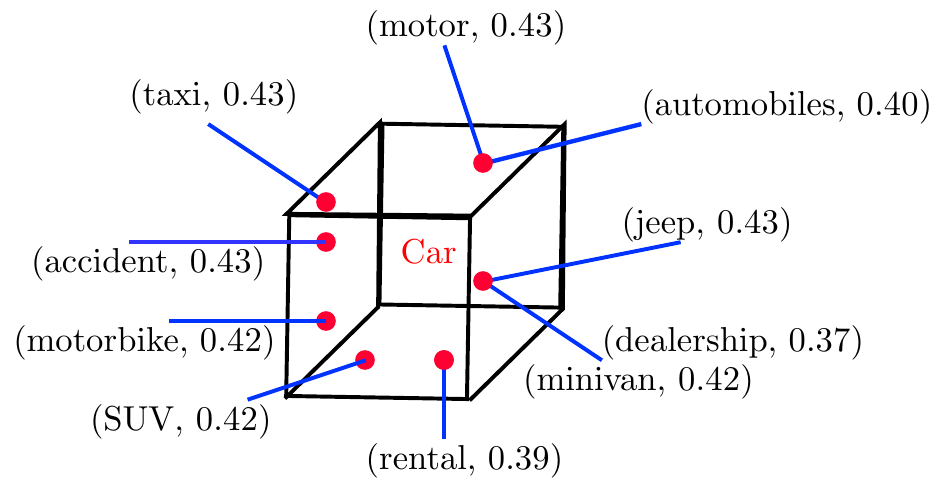}
  \vspace{-0.5cm}
\begin{center}
\end{center}
\endminipage
\vspace{0.1cm}
\minipage{0.333\textwidth}
  \includegraphics[width=\linewidth]{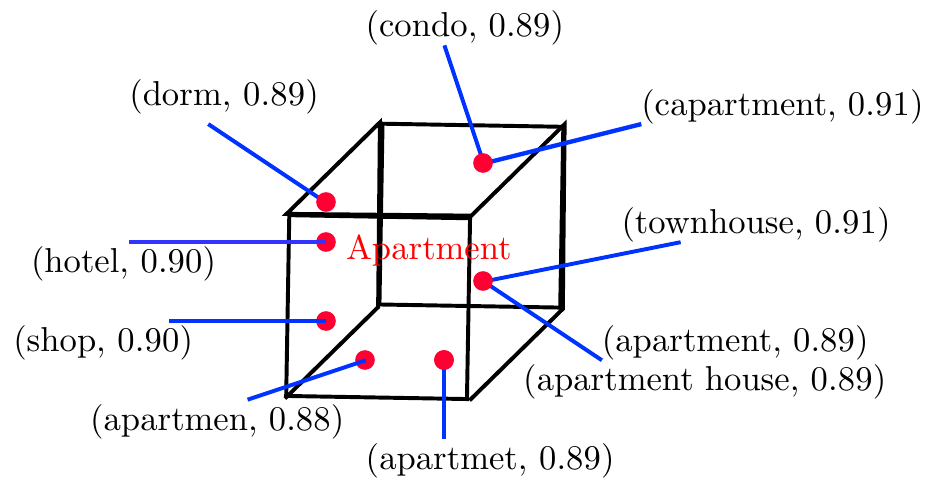}
 \vspace{-0.5cm}
\begin{center}

\end{center}
\endminipage\hfill
\minipage{0.333\textwidth}
  \includegraphics[width=\linewidth]{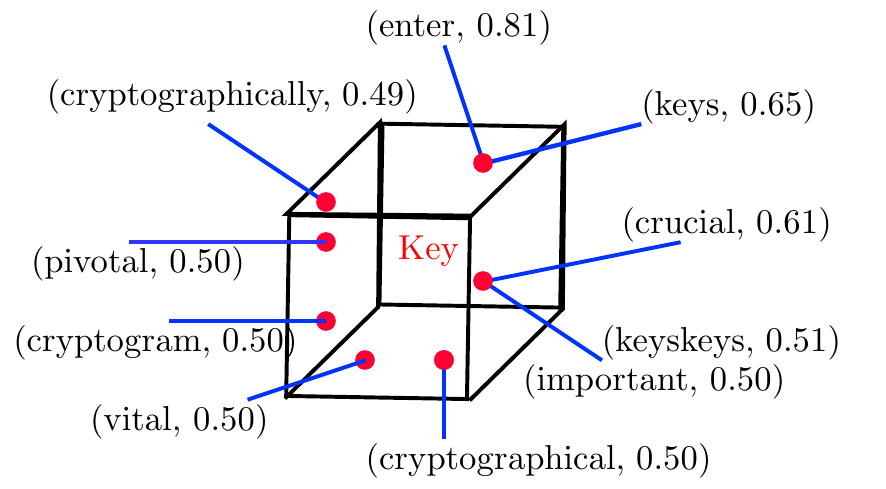}
 \vspace{-0.5cm}
\begin{center}
\textbf{(e) FastText}
\end{center}
\endminipage\hfill
\minipage{0.333\textwidth}%
  \includegraphics[width=\linewidth]{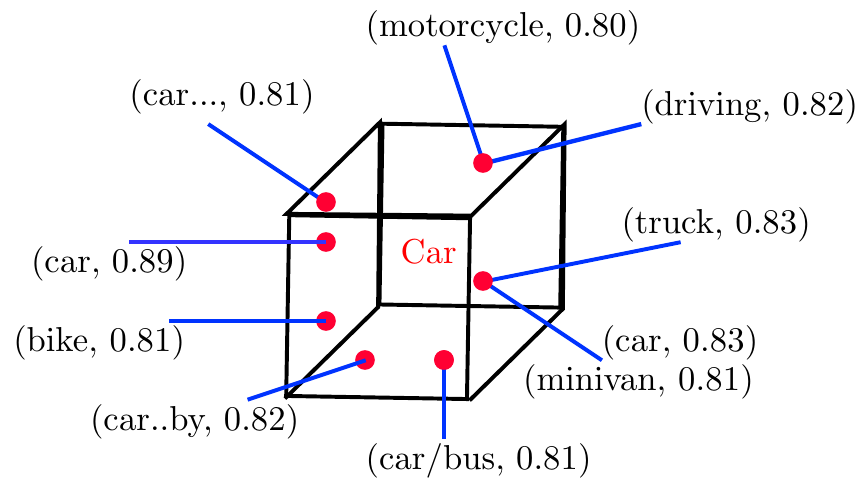}
  \vspace{-0.5cm}
\begin{center}
\end{center}
\endminipage 

\caption{Top 10 features for the words \textbf{\textit{Apartment}} (left), \textbf{\textit{Key}} (center) and \textbf{\textit{Car}} (right) generated from the five word embedding methods}
\label{fig:apartment_car_features}
\vspace{-0.2cm}
\end{figure*}

The key aspect that lies at the heart of many of the fMRI decoding studies is the establishment of an associative mapping of the linguistic representation of nouns or verbs and the corresponding brain activation patterns elicited when subjects viewed these lexical items. Mitchell's team designed a computational model to predict the brain responses using hand-crafted word vectors as input to map the correlation between word embeddings and brain activity involved in viewing the words~\cite{mitchell:predicting}. Since, Mitchell's 25 dimensional(dim) vector that uses fixed set of contextual dim (such as see, hear, eat etc) will face word sense disambiguity and our high dimensional word vectors would have better basis for sense disambiguation as they use co-occurrence frequencies from large corpora. For example, the lexical item ``Bank" has multiple semantic senses, such as the ``bank of a river" or a ``financial institution" based on the context. In fact, this forms motivation for our proposal of word embeddings in place of fixed context vectors. In recent times, linguistic representation of lexical items in computational linguistics is largely through a dense, low-dimensional and continuous vector called word-embedding~\cite{hinton:distributed,turney:frequency}. Common word embeddings are generated from large text corpora such as Wikipedia and statistics concerning the co-occurrence of words is estimated to build such embeddings~\cite{mikolov:distributed,bojanowski:enriching}. Some of the most popular word embedding models are Word2Vec~\cite{mikolov:distributed}, GloVe~\cite{pennington:glove} and Meta-Embeddings~\cite{yin:learning}. The recent popular approach FastText~\cite{bojanowski:enriching} is a fast and effective method to learn word representations and can be utilized for text classification. Since FastText embeddings are trained for understanding morphological variations and most of the syntactic analogies are morphology-based, FastText embeddings do significantly better on the syntactic analogies than on semantic tasks~\cite{bojanowski:enriching}. ~\cite{mikolov:distributed} introduced continuous Skip-gram model in Word2Vec that is an efficient method for learning high-quality distributed vector representations that capture a large number of precise syntactic and semantic word relationships. Global Vectors for word representations (GloVe)~\cite{pennington:glove} model combines the benefits of the Word2Vec skip-gram model when it comes to word analogy tasks, with the benefits of matrix factorization methods that can exploit global statistical information. In~\cite{yin:learning}, the idea of Meta-Embeddings has been proposed and has two benefits compared to individual embedding sets: enhancement of performance and improved coverage of the vocabulary. 

Recently, the success of deep learning based word representations has raised the question whether these models might be able to make association between brain activations and language. In~\cite{abnar2018experiential}, authors proposed a model that combines the experience based word representation model with the dependency based word2vec features. The resulting model yielded better accuracies. However, this paper does not discuss which are most predicted voxels in various brain regions for different word embedding models and also does not give results on brain activations corresponding to multiple senses of a word.  A recently published article~\cite{pereira2018toward} that gives a strong, independent support for our proposed approach of using word embedding representations for brain decoding models in place of carefully hand-crafted feature vectors. This paper aims at building a brain decoding system in which words and sentences are decoded from corresponding brain images. However, our approach addresses, in addition, the encoding problem where we try to build a system which learns associative mapping encoding words into corresponding fMRI images. Also, this paper uses ridge regression whereas we used Multi-layer feed forward neural network to learn the non-linear associative mapping between semantic features and brain activation responses.

In this paper, we propose a method to study the correlation between brain activity involved in viewing a word and corresponding word embedding (such as Word2Vec, GloVe, Meta-Embeddings, FastText and Mitchell's 25~\cite{mitchell:predicting}). To the best of our knowledge, this is the first time a comparative study is made of various existing, popular word embeddings for decoding brain activation. We propose a three-layer neural network architecture in which the input is a word embedding vector and the target output is the fMRI image depicting brain activation corresponding to the input word in line with the state-of-the-art approaches \cite{mitchell:predicting}.

The structure of the paper is as follows. In Section 2, we discuss the motivation towards using word embeddings. Section 3 describes the approach we are using to build the model, while Section 4 presents comparative results of various models along with the statistical significance of the results. In Section 5, we give the conclusions and future work.

\section{Motivation for using Word-Embeddings}

The word embeddings like Word2Vec, Glove etc., are known to capture the semantics of words based on the context as well as the co-occurrence of different words. We use these as features to capture the associative relationship between the meaning encoded in word embedding and the observed brain activation. So, Whenever the brain looks at a word, we assume that it tries to relate the word with some object/action, its properties, and other words with similar meaning. We consider the following example.

We observe the top 10 similar words for \textbf{Apartment}, \textbf{Key}, and \textbf{Car}. We obtain these similar words using different word embeddings which are given in Figure~\ref{fig:apartment_car_features}. In Word2Vec, the similar words are semantically similar to Apartment, key and Car. On the other hand, GloVe and Meta-Embeddings give not only semantically similar words but also related words like \textit{\{rental, parked, accident, insurance, etc.\}}  for \textbf{Car}, \textit{\{role, decisive, passwords, activation, leadership, etc.\}}  for \textbf{Key} and \textit{\{furnished, rent, renovated, etc.\}} for \textbf{Apartment}. These related words have the higher probability in Meta-Embeddings approach compared to those obtained with GloVe Embedding. These word embeddings are generated using just the text data without considering any brain activity specific features.

\begin{table*}[htbp]
\caption{Top 10 features for the word \textbf{Celery} generated from the six methods}
\footnotesize\setlength{\tabcolsep}{2.9pt}
\centering
\begin{tabular}{c| c | c | c | c | c | c } \hline \hline
(1) & (2) & (3) &(4)&(5)  & (6) & (7)\\ \hline \hline
broccoli 0.71&eat 0.35&carrots 0.16 & carrots 0.24&eat 0.19& eat 0.837 & cabbage 0.74\\ 
bellpeppers 0.69 &taste 0.24&onions 0.16 & cabbage 0.33 &taste 0.18&taste 0.346 & carrots 0.74  \\ 
parsley 0.69&fill 0.051&parsley 0.18 & cauliflower 0.35 &fill 0.012& fill 0.315 &onions 0.73\\ 
cilantro 0.68 &see 0.063&broccoli 0.20 & onion 0.35 &see 0.07&see 0.243 &spinach 0.73\\ 
cabbage 0.68&clean 0.054&garlic 0.20 & parsley 0.38 &clean 0.018 &clean 0.115 & garlic 0.72\\ 
cauliflower 0.67  &open 0.042& cabbage 0.21 &broccoli 0.38 &open 0.08&open 0.060 & tomato 0.70\\
tomato 0.67 &smell 0.189&carrot 0.21 & garlic 0.38 &smell 0.026&smell 0.059 &potatoes 0.70 \\  
lettuce 0.67&touch 0.061&spinach 0.22& potatoes 0.40&touch 0.019&touch 0.029 & parsnips 0.69\\ 
cherry 0.66 &say 0.094&cauliflower 0.22 & turnips 0.40 &say 0.092&say 0.016 & sweetroot 0.69\\ 
Brussels 0.66 &hear 0.021&asparagus 0.23 & lettuce 0.41 &hear 0.032&hear 0.000 & lemongrass 0.69\\
\hline \hline 
\multicolumn{7}{l}{(1): Word2Vec(Top 10), (2): Word2Vec Similarity (with Mitchell's 25 words), (3): GloVe(Top 10)}\\
\multicolumn{7}{l}{(4): Meta-Embeddings (Top 10), (5): Meta-Embeddings Similarity (with Mitchell's 25 words) (Top 10)}\\
\multicolumn{7}{l}{(6): Mitchell's 25 (Top 10), (7): FastText (Top 10)}
\end{tabular}
\label{some_features}
\end{table*}

On the other hand, Mitchell's feature vectors would be, by design, related to stimulus-modality-specific brain regions, as the learning model associates sensory features that have large weights with dominant evoked responses in related sensory cortical areas. The word embedding methods (Word2Vec, GloVe, and Meta-Embeddings) encode the meaning in terms of co-occurrence frequencies of other words in the corpus and thus may not relate to various modules of the brain the way Mitchell's hand-crafted features are designed.

It is interesting to understand the closeness of various word embeddings with Michell's 25.  Table~\ref{some_features} describes similar words for ``celery" based on various word embeddings as well as Mitchell's 25. As word embeddings and Mitchell's operate on different dimensions, we checked if they have similar underlying semantics. We estimated similarity scores for embedding vector for celery with embedding vectors for various feature words used in Mitchell's. Table~\ref{some_features} shows that the resulting score vector is quite similar, pointing out that the underlying similarity of semantics between vector-based encoding and Mitchell's. In this way, even these methods seem to capture the meaning in a way similar to Mitchell's scheme and perhaps might learn to elicit appropriate brain activation. 

\section{Proposed Approach}
\vspace{-0.4cm}
\label{method1}
\begin{figure}[h]
\begin{center}
\includegraphics[width=0.85\linewidth]{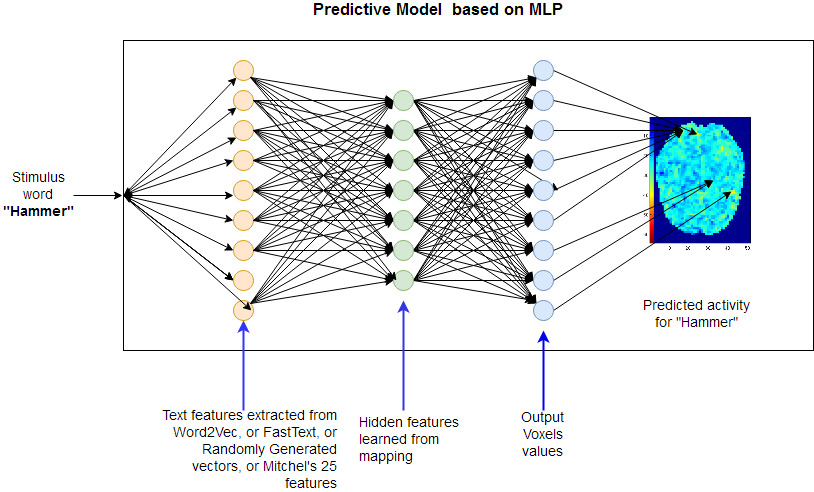}
\caption{ \textbf{3-layer neural network architecture for decoding fMRI brain activation}}
\label{fig:MLP}
\end{center}
\end{figure}
In this paper, we use a 3-layer neural network architecture as shown in Figure~\ref{fig:MLP} to build a trainable computational model that predicts the neural activation for any given stimulus word \textbf{(w)}.
Given a random stimulus word \textbf{(w)}, we provide semantic features associated with \textbf{(w)} as input (generated from one of the six different methods, namely, Word2Vec, GloVe, Meta-Embeddings, FastText, Randomly-generated, and Mitchell's 25 \cite{mitchell:predicting}). 
The second step involves hidden layer representation and is accomplished via $N$ hidden neurons in the hidden layer. Hidden neurons are fully connected to the input layer and the connection weights are learned through an adaptation process. The third step predicts the neural fMRI activation at every voxel location in the brain as a weighted sum of neural activations contributed by each of the hidden layer neurons. More precisely, the predicted activation $z_{v}$ at voxel $v$ in the brain for word $w$ is given by
\begin{align*}
z_{v}&=\sum_{j=1}^{N}c_{vj}f(net_{j})+c_{j0}\\
f(net_{j})&=tanh(\sum_{i=1}^{M}c_{ij}x_{i}+c_{i0})
\end{align*}
where, $f(net_{j})$ is the value of the $j^{th}$ hidden neuron for word $w$, $N$ is the number of hidden neurons present in the model, and $c_{vj}$ is a learned coefficient that specifies the degree to which the $j^{th}$ intermediate semantic feature activates a voxel in the output layer. 

\section{Experimental Results and Observations}
\label{headings}
In this section, we describe the details of experiments conducted and the use of various word embeddings and observations thereof. We first describe the datasets used for our study.
\subsection{fMRI Dataset Description}
We used CMU fMRI data\footnote{Available at \url{http://www.cs.cmu.edu/~fmri/science2008/data.html}} of nine healthy subjects. These nine healthy subjects viewed 60 different word-picture pairs six times each. The 60 arbitrary stimuli included five items from each of the 12 semantic categories (animals, body parts, building parts, buildings, furniture, clothing, insects, kitchen items, tools, vegetables, vehicles, other man-made items). For each stimulus, we computed a mean fMRI image over its six repetitions and the mean of all 60 of these stimuli was then subtracted to get the final representation image.

\begin{figure*}[t]
\minipage{0.14\textwidth}
  \includegraphics[width=\linewidth]{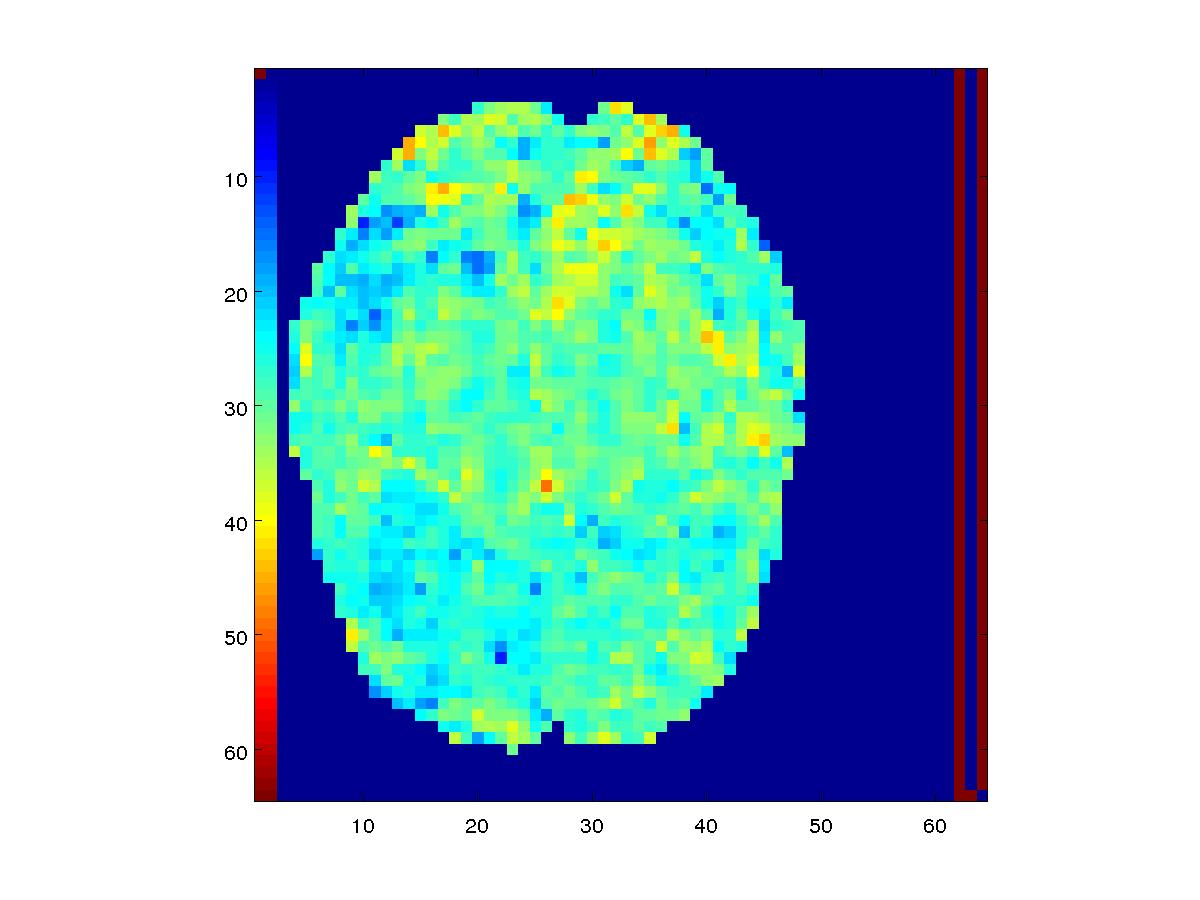}  
 \vspace{-0.5cm}
\begin{center}
\textbf{(a) Original}
\end{center}
\endminipage\hfill
\minipage{0.155\textwidth}
  \includegraphics[width=\linewidth]{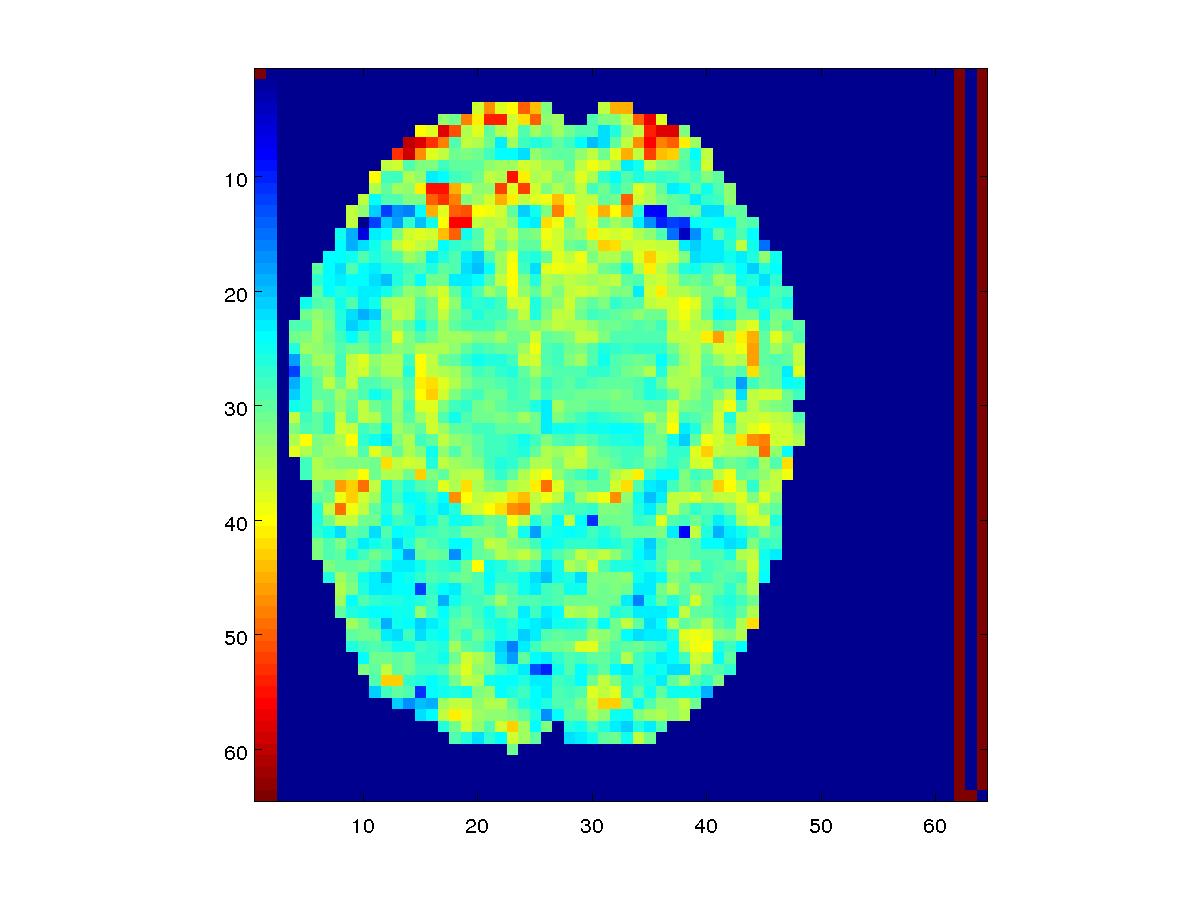}
  \vspace{-0.5cm}
\begin{center}
\textbf{(b) Word2Vec}
\end{center}
\endminipage\hfill
\minipage{0.145\textwidth}
  \includegraphics[width=\linewidth]{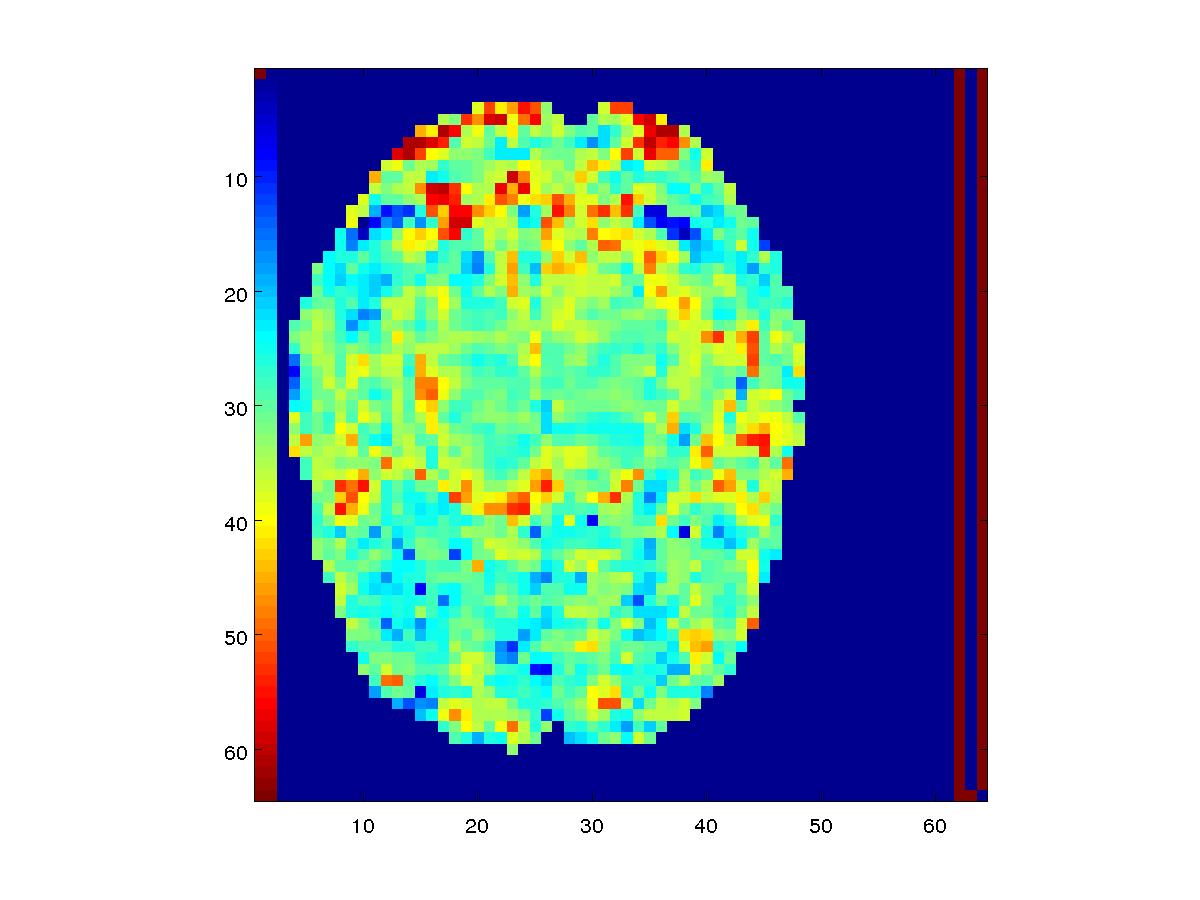}
 \vspace{-0.5cm}
\begin{center}
\textbf{(c) Mitchell's}
\end{center}
\endminipage\hfill
\minipage{0.14\textwidth}
  \includegraphics[width=\linewidth]{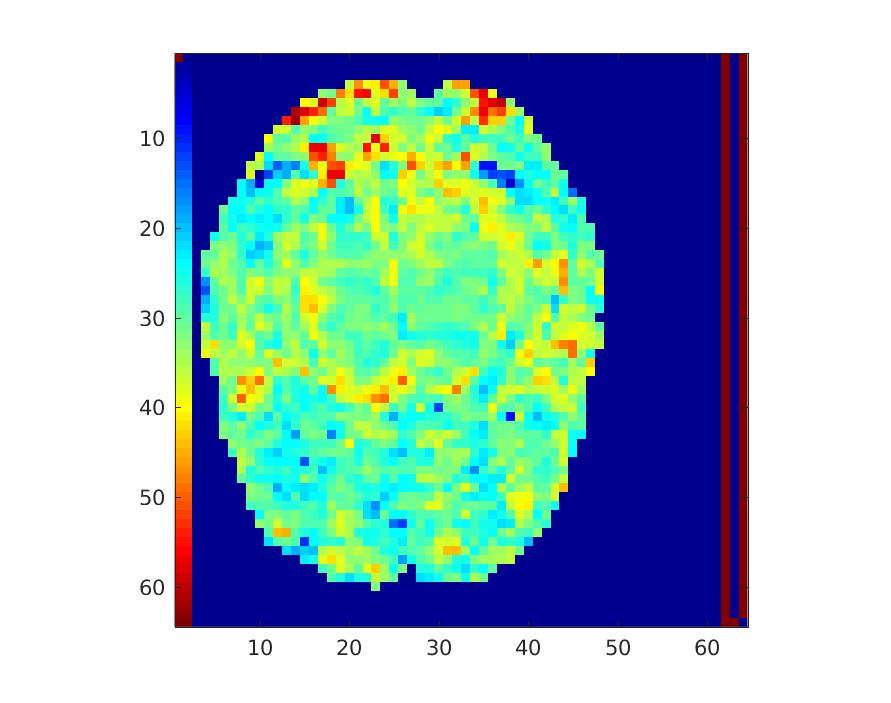}
  \vspace{-0.5cm}
\begin{center}
\textbf{(d) GloVe}
\end{center}
\endminipage\hfill
\minipage{0.14\textwidth}
  \includegraphics[width=\linewidth]{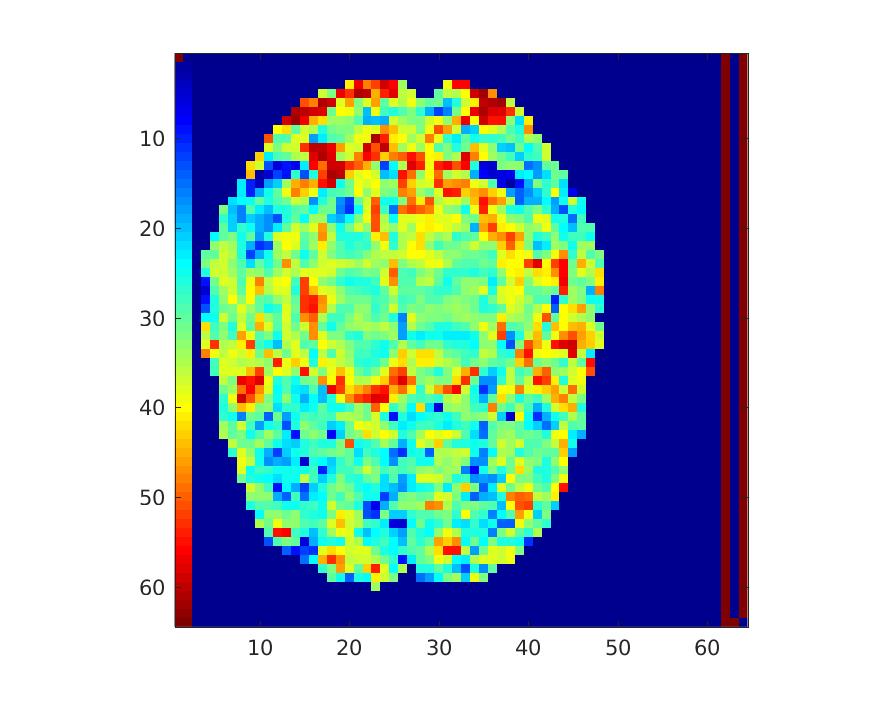}
 \vspace{-0.5cm}
\begin{center}
\textbf{(e) Meta-Embeddings}
\end{center}
\endminipage\hfill
\minipage{0.14\textwidth}%
  \includegraphics[width=\linewidth]{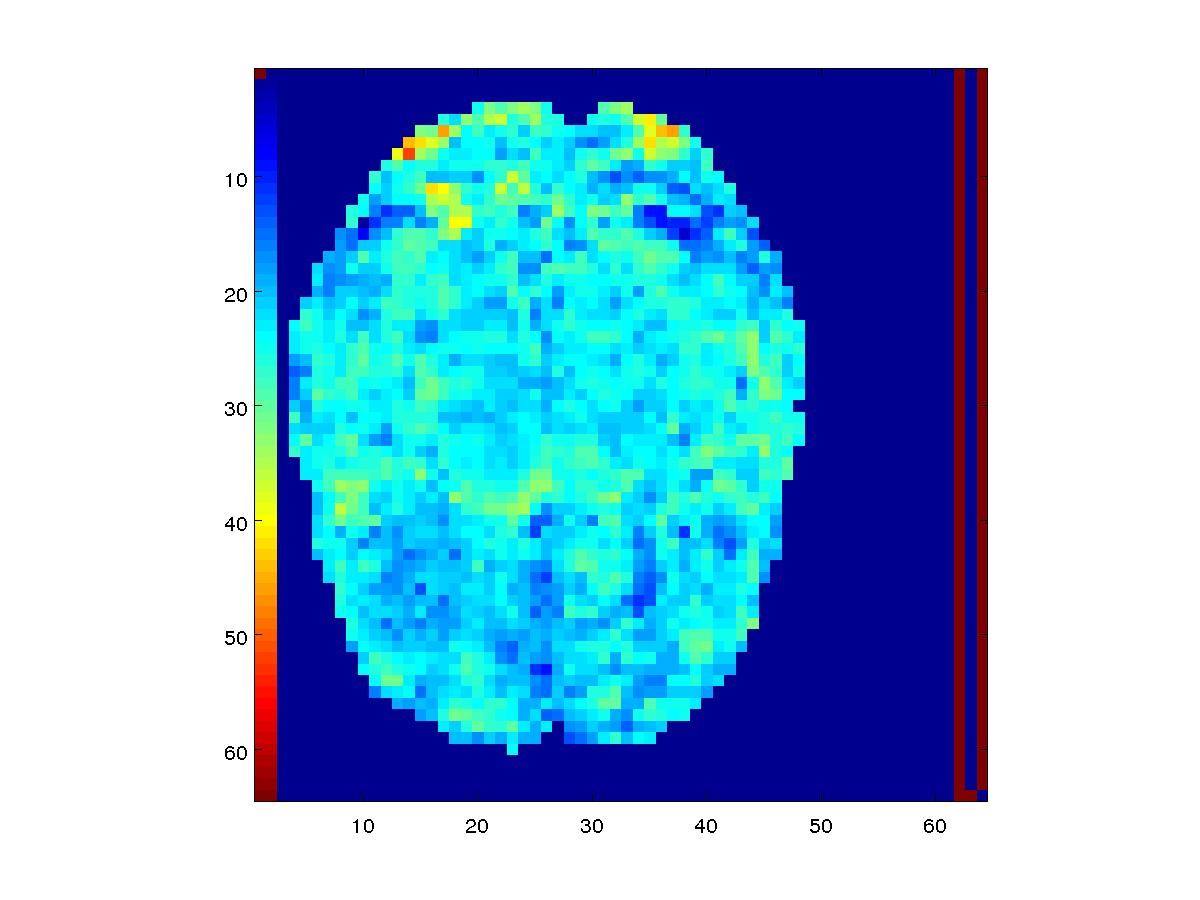}
  \vspace{-0.5cm}
\begin{center}
\textbf{(f) Random}
\end{center}
\endminipage\hfill
\minipage{0.14\textwidth}%
  \includegraphics[width=\linewidth]{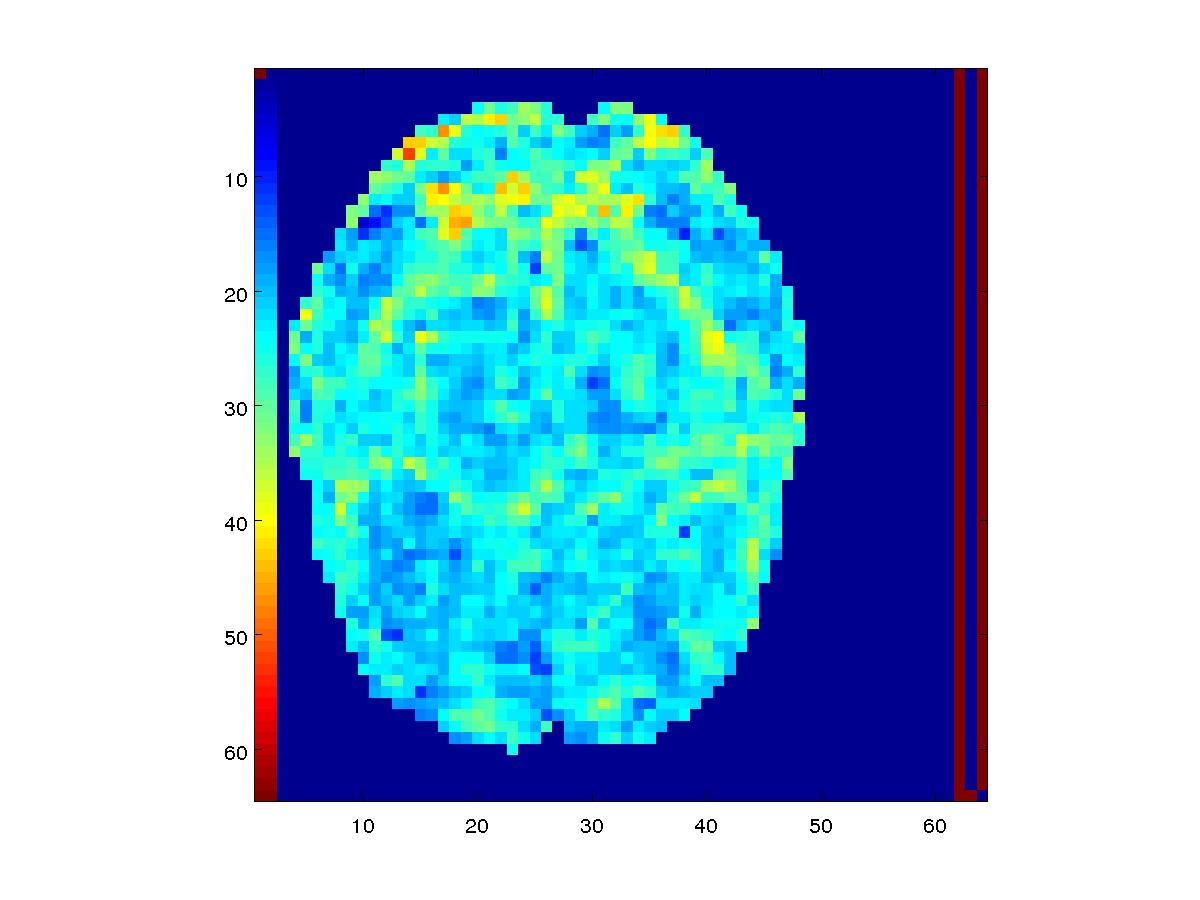}
  \vspace{-0.5cm}
\begin{center}
\textbf{(g) FastText}
\end{center}
\endminipage 
\caption{Predicting fMRI images
for given stimulus word ``bell"}
\minipage{0.14\textwidth}
  \includegraphics[width=\linewidth]{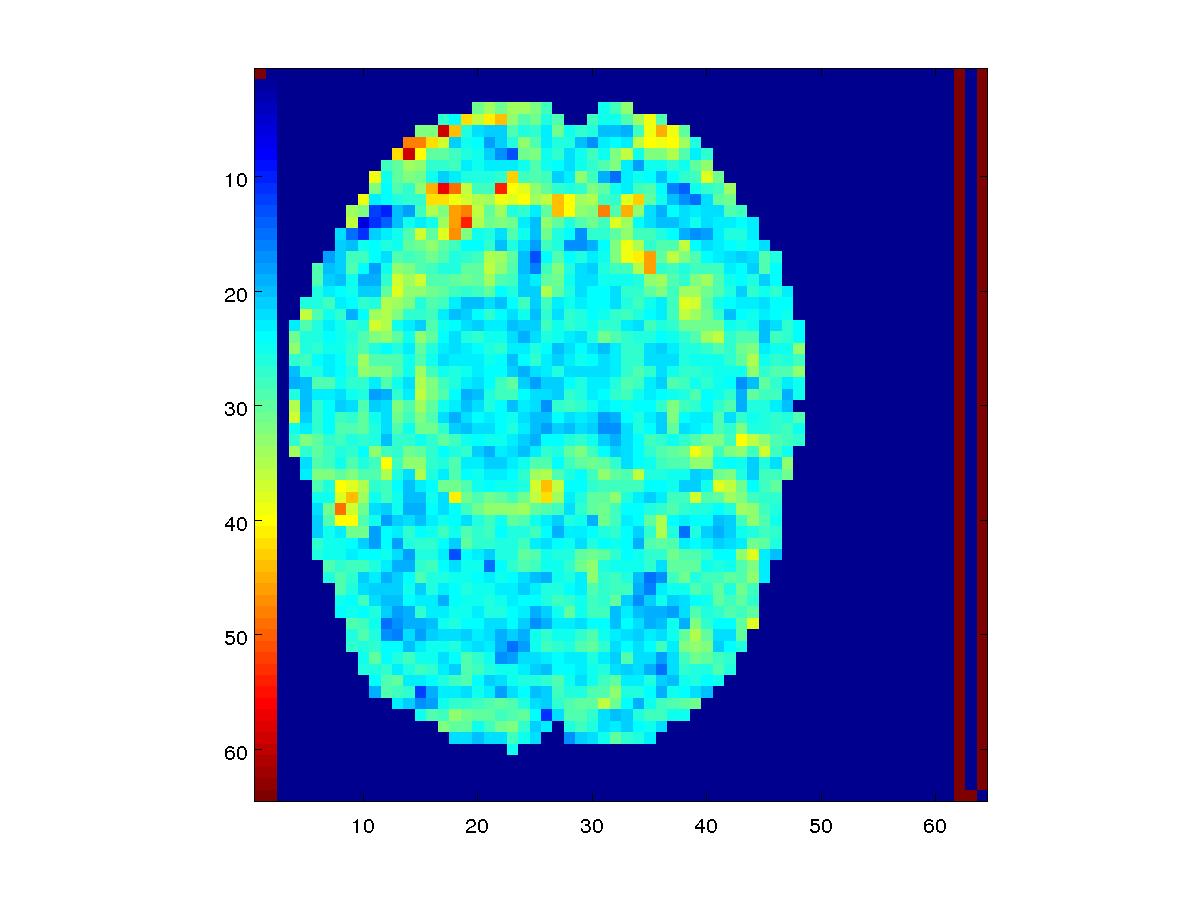}
  \vspace{-0.5cm}
\begin{center}
\textbf{(a) Original}
\end{center}
\endminipage\hfill
\minipage{0.155\textwidth}
  \includegraphics[width=\linewidth]{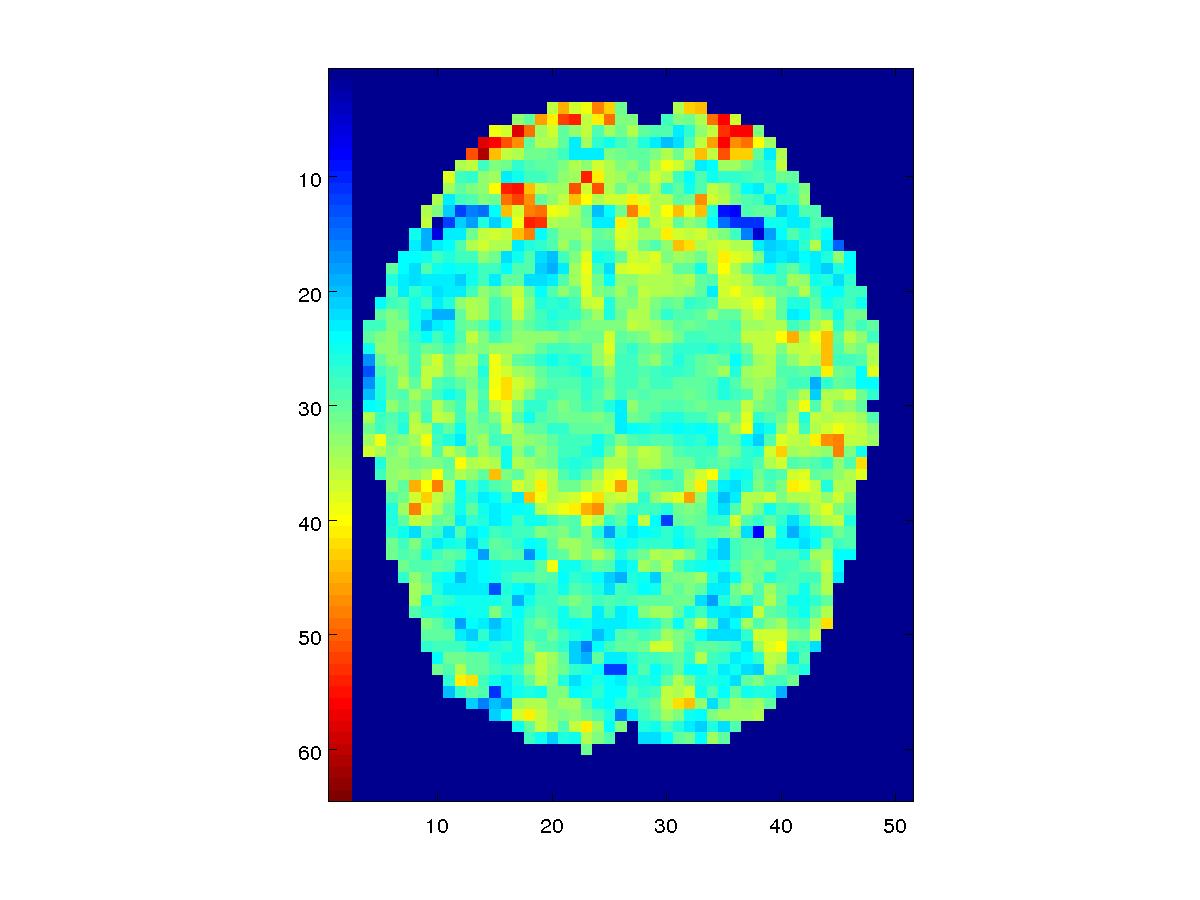}
  \vspace{-0.5cm}
\begin{center}
\textbf{(b) Word2Vec}
\end{center}
\endminipage\hfill
\minipage{0.145\textwidth}
  \includegraphics[width=\linewidth]{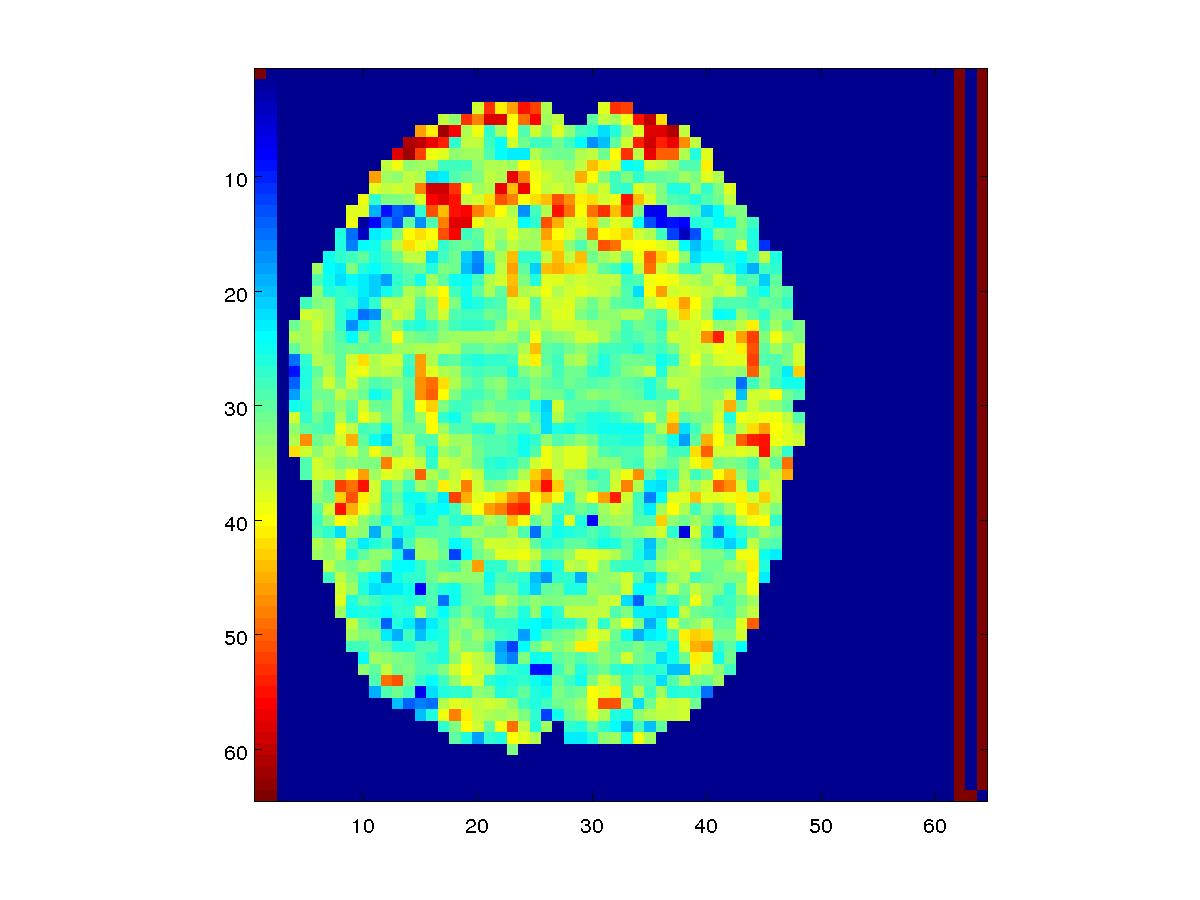}
  \vspace{-0.5cm}
\begin{center}
\textbf{(c) Mitchell's}
\end{center}
\endminipage\hfill
\minipage{0.14\textwidth}
  \includegraphics[width=\linewidth]{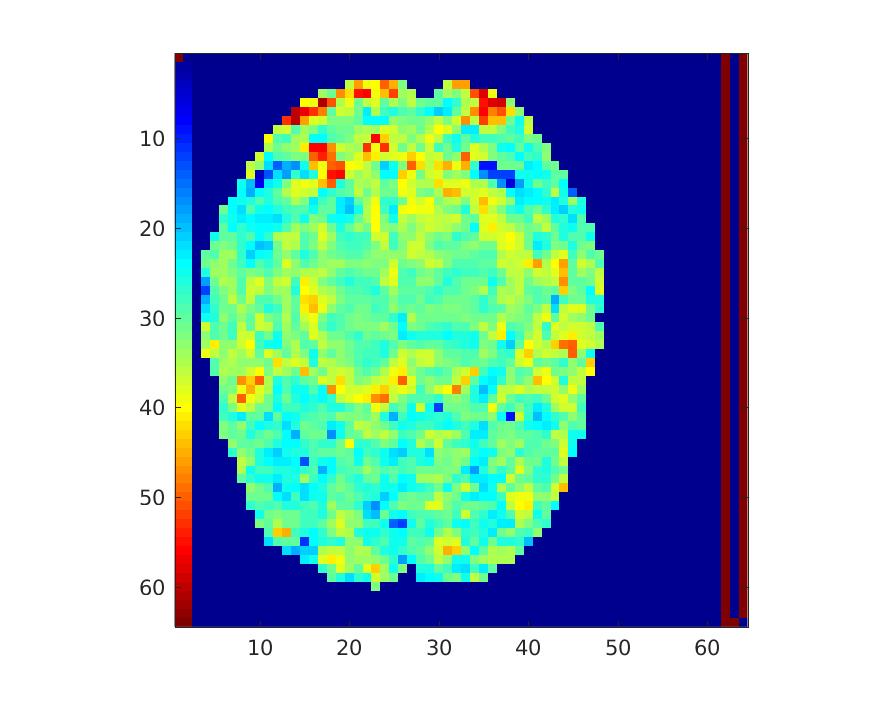}
  \vspace{-0.5cm}
\begin{center}
\textbf{(d) GloVe}
\end{center}
\endminipage\hfill
\minipage{0.14\textwidth}
  \includegraphics[width=\linewidth]{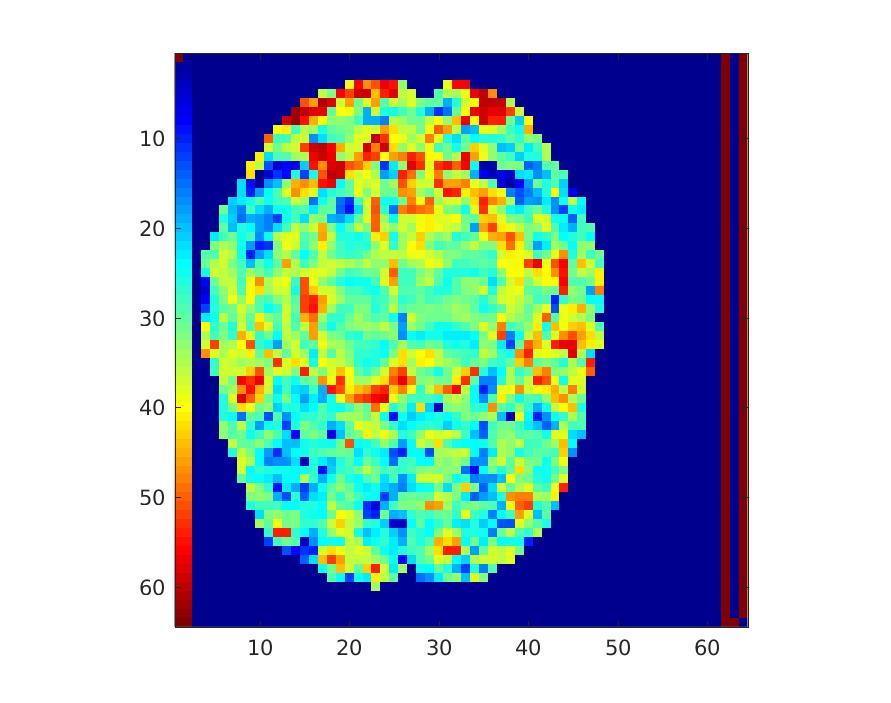}
  \vspace{-0.5cm}
\begin{center}
\textbf{(e) Meta-Embeddings}
\end{center}
\endminipage\hfill
\minipage{0.14\textwidth}%
  \includegraphics[width=\linewidth]{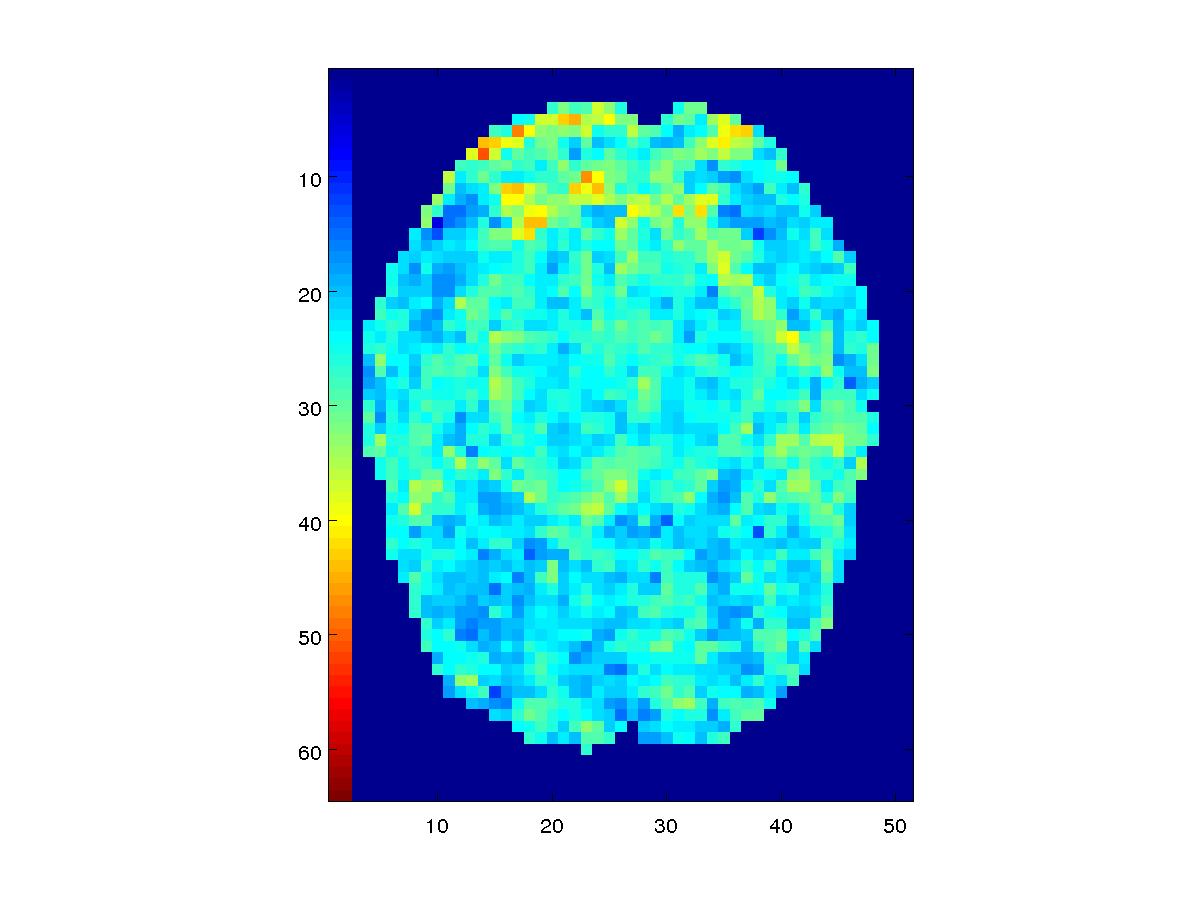}
  \vspace{-0.5cm}
\begin{center}
\textbf{(f) Random}
\end{center}
\endminipage\hfill
\minipage{0.14\textwidth}%
  \includegraphics[width=\linewidth]{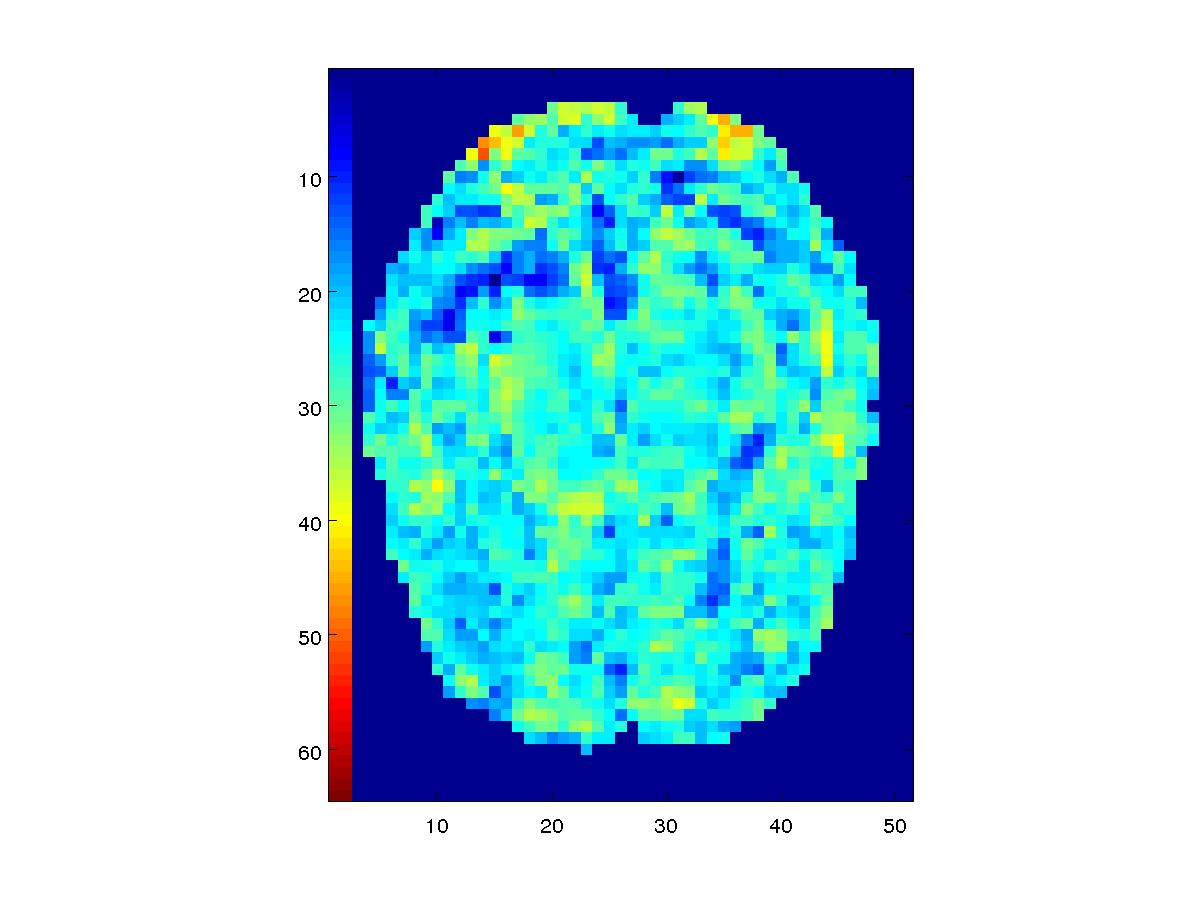} 
  \vspace{-0.5cm}
\begin{center}
\textbf{(g) FastText}
\end{center}
\label{fig:awesome_image2}
\endminipage 
\caption{Predicting fMRI images
for given stimulus word ``arm"}
\minipage{0.14\textwidth}
  \includegraphics[width=\linewidth]{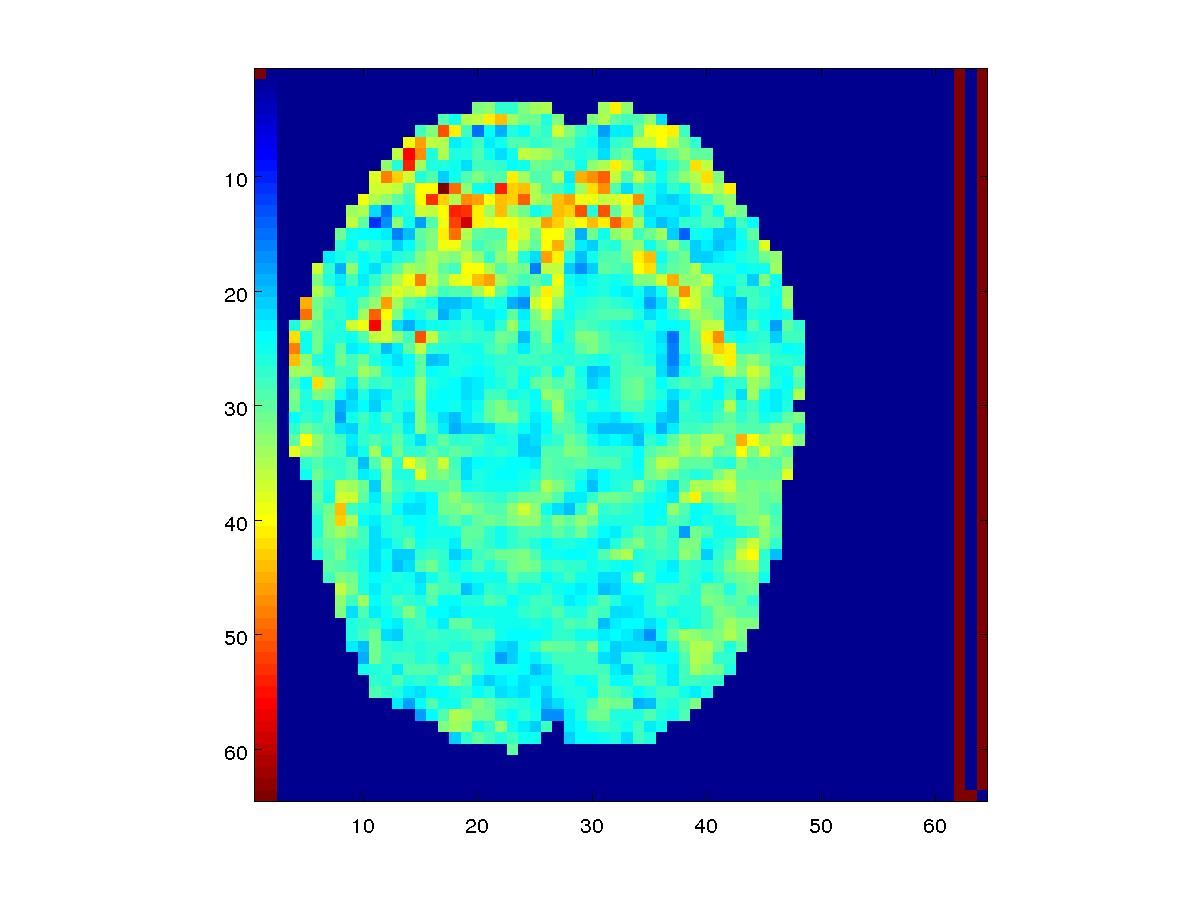}
  \label{fig:awesome_image3}
  \vspace{-0.5cm}
\begin{center}
\textbf{(a) Original}
\end{center}
\endminipage\hfill
\minipage{0.155\textwidth}
  \includegraphics[width=\linewidth]{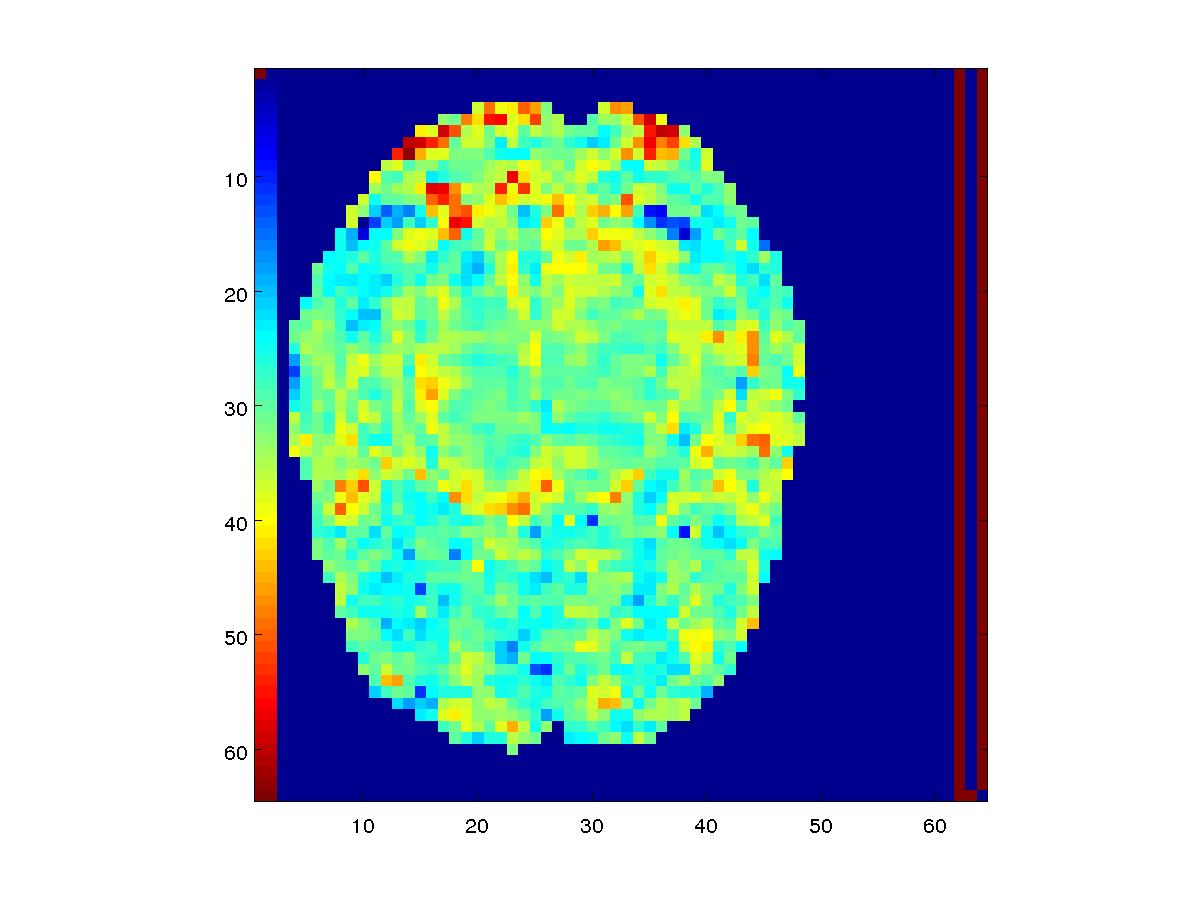}
  \vspace{-0.5cm}
\begin{center}
\textbf{(b) Word2Vec}
\end{center}
\endminipage\hfill
\minipage{0.145\textwidth}
  \includegraphics[width=\linewidth]{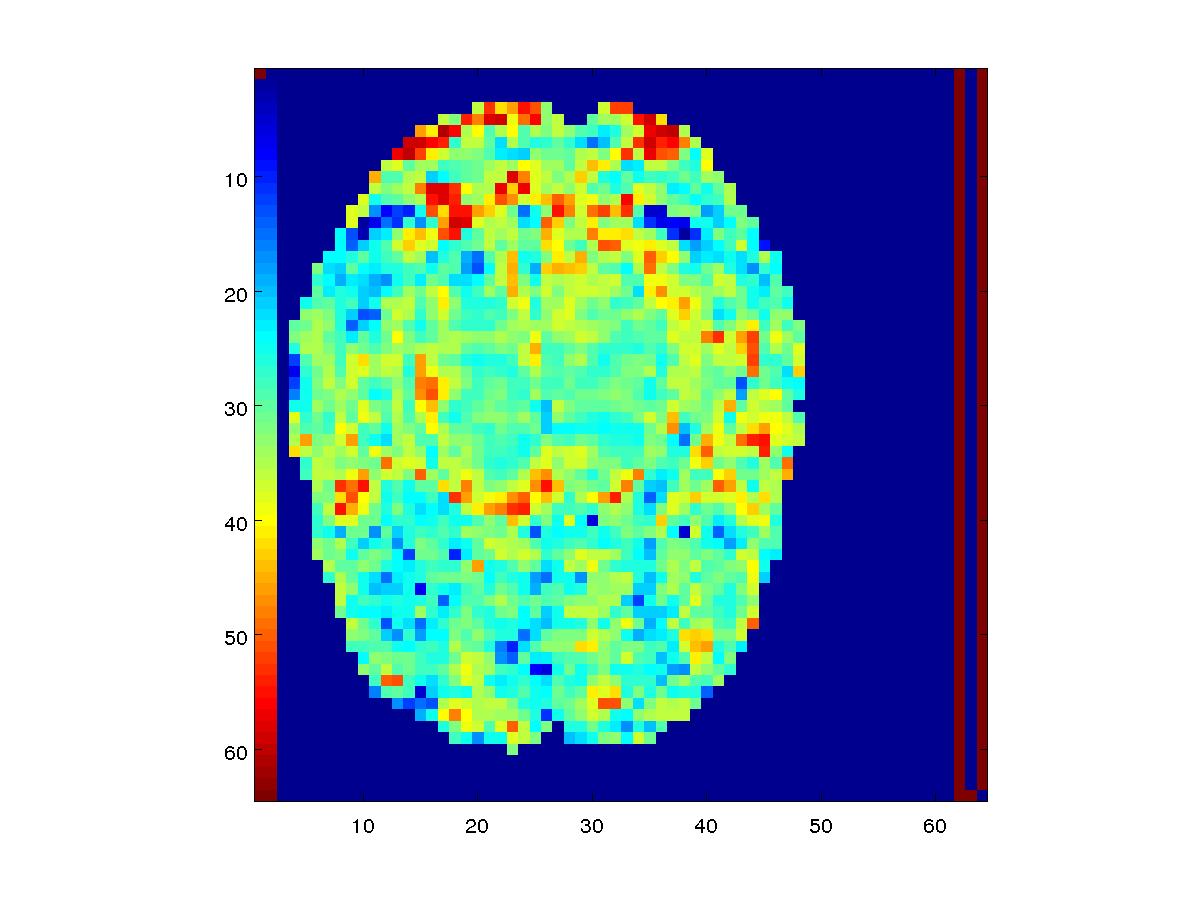}
  \vspace{-0.5cm}
\begin{center}
\textbf{(c) Mitchell's}
\end{center}
\endminipage\hfill
\minipage{0.14\textwidth}
  \includegraphics[width=\linewidth]{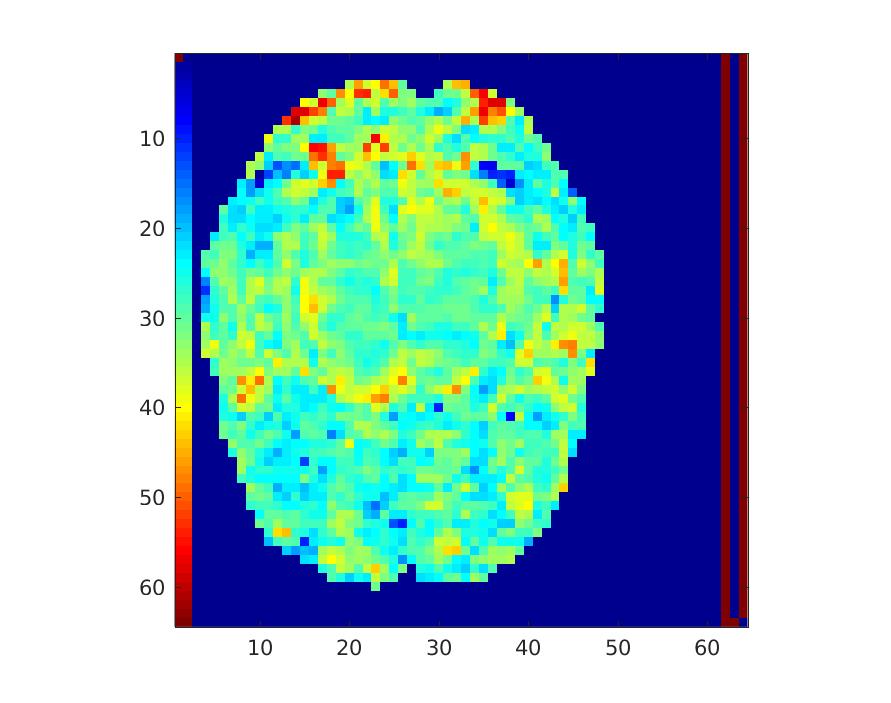}
  \vspace{-0.5cm}
\begin{center}
\textbf{(d) GloVe}
\end{center}
\endminipage\hfill
\minipage{0.14\textwidth}%
  \includegraphics[width=\linewidth]{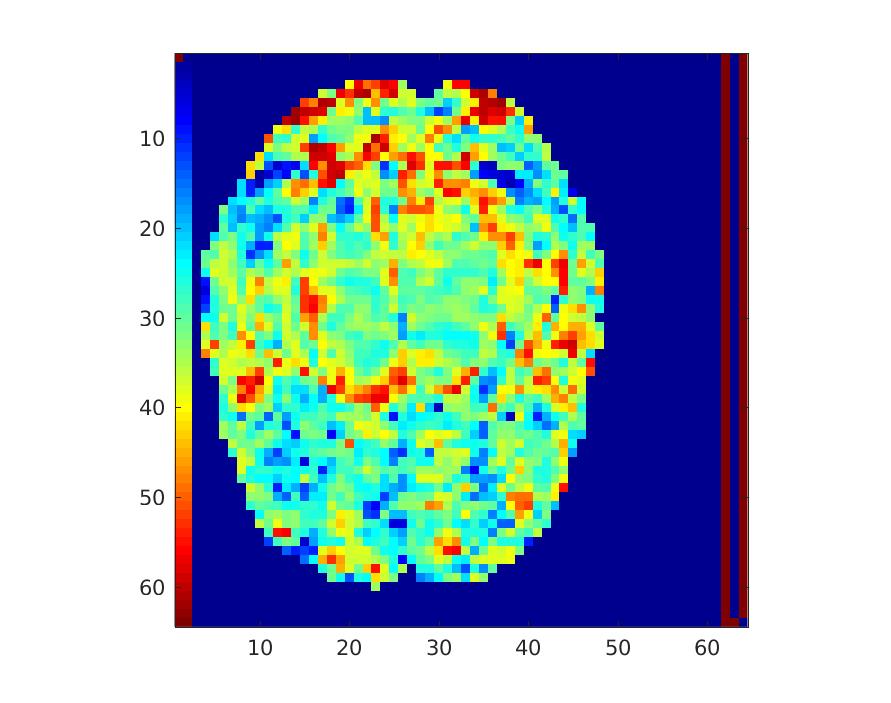}
  \vspace{-0.5cm}
\begin{center}
\textbf{(e) Meta-Embeddings}
\end{center}
\endminipage\hfill
\minipage{0.14\textwidth}%
  \includegraphics[width=\linewidth]{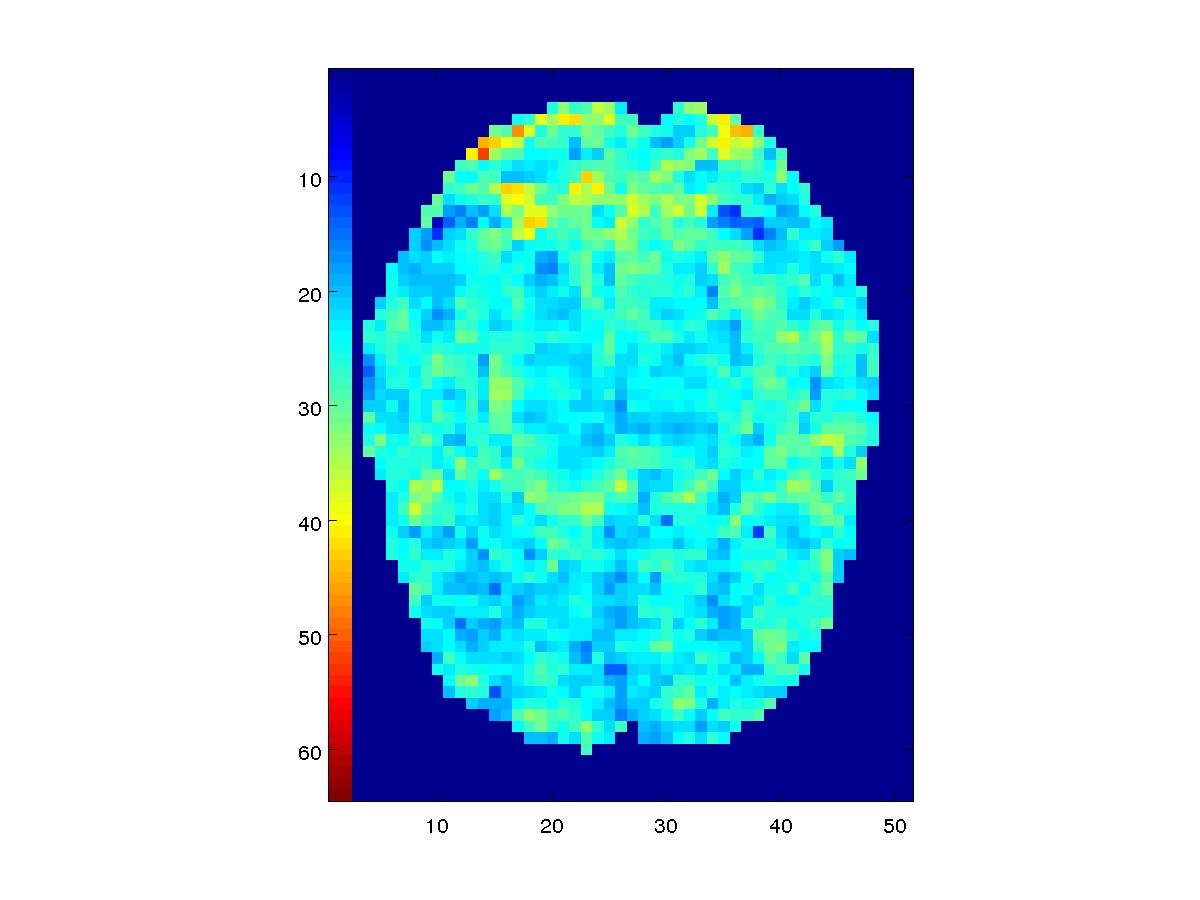}
  \vspace{-0.5cm}
\begin{center}
\textbf{(f) Random}
\end{center}
\endminipage\hfill
\minipage{0.14\textwidth}%
  \includegraphics[width=\linewidth]{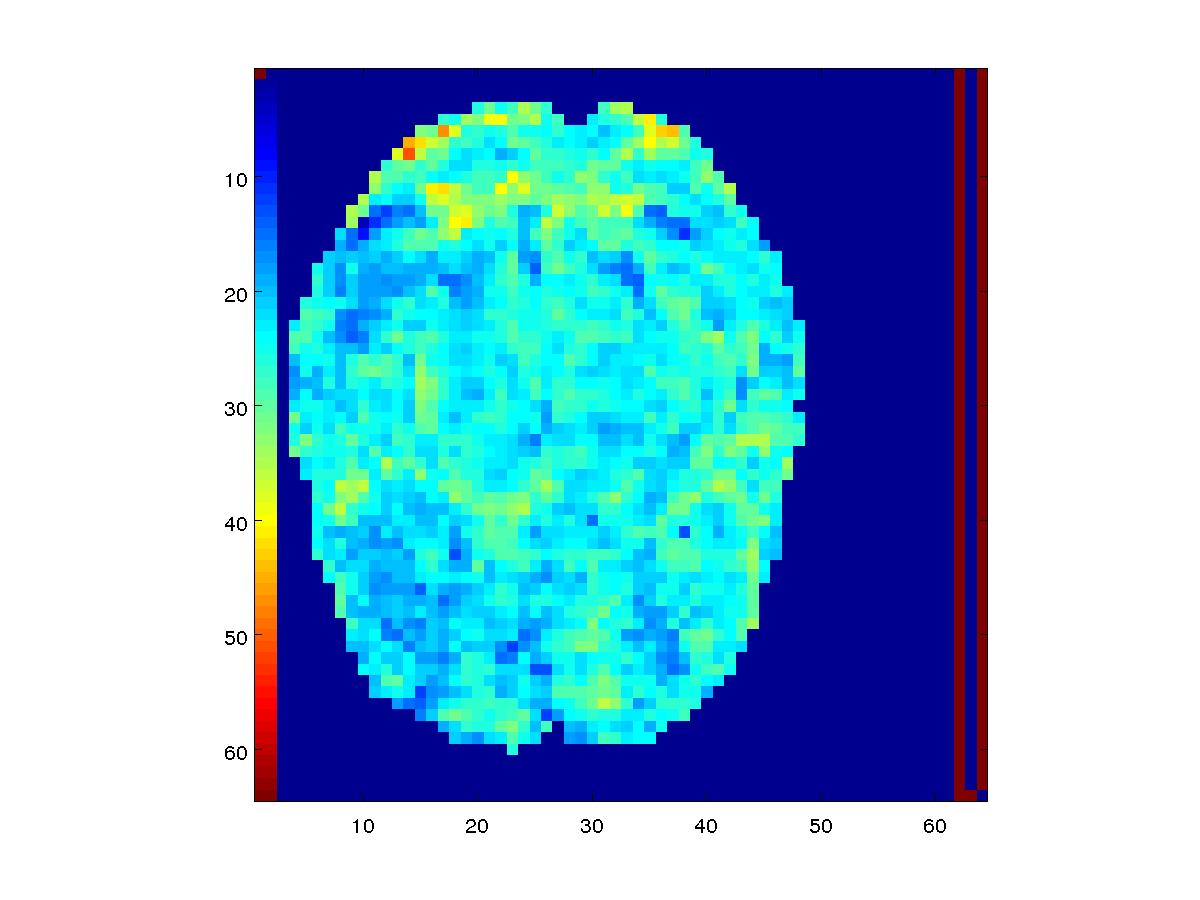}
  \vspace{-0.5cm}
\begin{center}
\textbf{(g) FastText}
\end{center}
\endminipage
\caption{Predicting fMRI images
for given stimulus word ``bee"}
\label{fig:awesome_image1}
\end{figure*}

\subsection{Architecture used and Training strategy}
\begin{table}[h]
\caption{3-layer neural network parameter setting}
\footnotesize\setlength{\tabcolsep}{2.9pt}
\centering
\begin{tabular}{l||c } \hline \hline
Parameters&Values\\ \hline \hline
Hidden layer size & 100\\ 
Optimizer& Adam \\
Activation & Tanh \\
Momentum, Learning rate& 0.9, 0.001 \\
\hline \hline 
\end{tabular}
\label{parameters}
\end{table}
The 3-layer neural network had 100 nodes in the hidden layer. Table~\ref{parameters} describes the other parameter settings for proposed model. At the input layer, we use semantic features of stimulus word. These semantic features could be any one of Word2Vec, GloVe, Meta-Embeddings, FastText, Randomly generated vectors, or Mitchell's 25 features. The reason behind using random features is to set a baseline control study. 
We trained separate computational models for each of the 9 participants using all the four input encoding methods. Each trained model was evaluated by means of a ``leave-one-out" cross-validation approach in which the model was repeatedly trained with 59 of the 60 available word stimuli and associated fMRI images. Each trained model was then tested by requiring it to predict the fMRI image for the one ``held-out" word. 

\subsection{Statistical Analysis of Predicted fMRI Images}
Figures \ref{fig:awesome_image2}, \ref{fig:awesome_image3} and \ref{fig:awesome_image1} compare the ground truth fMRI image and the corresponding predicted fMRI images using all the six methods for the words ``bell", ``arm" and ``bee". It can be observed from the Figures \ref{fig:awesome_image2}, \ref{fig:awesome_image3} and \ref{fig:awesome_image1} that the predicted fMRI images corresponding to Word2Vec, GloVe, and Mitchell's features look visually similar to the actual fMRI image obtained during the empirical experiment, whereas Random and FastText results differ significantly. 

The predicted fMRI images when Meta-Embeddings are used have more robust activation compared to that of the original fMRI images. From this we can infer that Meta-Embeddings which use multiple data sources, not only covers semantically similar words but also gets closer to how the brain seems to represent.  However, the activation regions seem largely similar in all the approaches except that of approaches using Random and FastText embeddings.
\vspace{-0.3cm}
\begin{table*}[!htb]
\caption{Statistical significance (one way ANOVA test) among the six methods reported individually per subject}
\footnotesize\setlength{\tabcolsep}{2.9pt}
\centering
\begin{tabular}{l|c c c c c c c c } \hline \hline
\#Subject&(1)&(2)&(3)& (4) &(5)&(6)&F statistic&p-value\\ \hline \hline
Subj-1& 0.2867& 0.5562 & 0.5587 & 0.5561 & -0.05600 & -0.0078& 17.6136 & 1.332e-15*\\ 
Subj-2& 0.2963 & 0.3169 & 0.3194 & 0.3064 & -0.0600 & -0.0089 & 16.1014 & 2.620e-14* \\ 
Subj-3& 0.2963& 0.2924 & 0.2972 & 0.2911 &-0.0600 & -0.0089& 13.1922 & 8.552e-12*\\ 
Subj-4& 0.4327 & 0.4273 & 0.4319 & 0.4253 & 0.3208 & 0.3435 & 7.6373 & 7.840e-07* \\ 
Subj-5& 0.1918& 0.1800 & 0.1883 & 0.1805 & -0.2231 & -0.5236& 29.7585 & 1.110e-16*\\ 
Subj-6& -0.8066  & -0.8213 & -0.8008 & -0.7797 & -1.2333 & -1.4631 & 1.4862 & 0.1935 \\
Subj-7& 0.2015 & 0.1896 &0.1961 & 0.1924 & -0.1820 & -0.1564 & 13.5018& 3.677e-08* \\  
Subj-8& 0.2270& 0.2200 & 0.2280 & 0.2213 & -0.1469 & -0.1710& 29.7879& 1.110e-16*\\ 
Subj-9& 0.1816 & 0.1751 & 0.1778 & 0.1735 & -0.3220 & -0.2670 & 15.0325 & 5.497e-09* \\ 
\hline \hline 
\multicolumn{9}{l}{(1): Word2vec, (2): Mitchell's 25, (3): Glove, (4): Meta-Embeddings, (5): Random, (6): FastText} \\
\multicolumn{6}{l}{*p$<$0.05}
\end{tabular}
\label{results}
\end{table*}

\vspace{-1.5cm}

\begin{table*}[!htb]
\caption{\emph{Post-hoc} multiple comparison of the six embedding schemes}
\label{results1}
\footnotesize\setlength{\tabcolsep}{2.9pt}
\centering
\begin{tabular}{l|c c c c c c c c c } \hline \hline
&\multicolumn{9}{ c }{\textbf{Subjects (significance)}} \\ 
Post-hoc&Subj-1&Subj-2&Subj-3&Subj-4&Subj-5&Subj-6&Subj-7&Subj-8&Subj-9\\ \hline \hline
(1) vs (2)& 0.001**& 0.8995 & 0.8995 & 0.8995& 0.8995& 0.8995& 0.89947 &0.89947& 0.89947\\ 
(1) vs (3)& 0.001**& 0.8995 & 0.8995 & 0.8995& 0.8995& 0.8995& 0.89947 &0.89947& 0.89947\\
(1) vs (4)& 0.001**& 0.8995 & 0.8995 & 0.8995& 0.8995& 0.8995& 0.89947 &0.89947& 0.89947\\
(1) vs (5)& 0.001**& 0.001** & 0.001** & 0.001**& 0.001**& 0.3799& 0.001** &0.001**& 0.001**\\
(1) vs (6)& 0.001**& 0.001** & 0.001** & 0.0087**& 0.001**& 0.7800& 0.001** &0.001**& 0.001**\\
\textbf{(2) vs (3)}& 0.8995 & 0.8995 & 0.8995 & 0.8995 & 0.8995 & 0.8995 & 0.8995 & 0.8995 & 0.8995\\ 
\textbf{(2) vs (4)}& 0.8995 & 0.8995 & 0.8995 & 0.8995 & 0.8995& 0.8995 & 0.8995 & 0.8995 & 0.8995\\  
(2) vs (5)& 0.4777& 0.001** & 0.001** & 0.001**& 0.001**& 0.4069& 0.0010** &0.001**& 0.001**\\ 
(2) vs (6)& 0.7998 & 0.001** & 0.001** & 0.0172* & 0.001** & 0.8050 & 0.001** & 0.001** & 0.001**\\ 
\textbf{(4) vs (3)} & 0.8995& 0.8995 & 0.8995 & 0.8995& 0.8995& 0.8995& 0.8995 &0.8995& 0.8995\\
(4) vs (5)& 0.4789& 0.001** & 0.001** & 0.001**& 0.001**& 0.3325& 0.001** &0.001**& 0.001**\\
(4) vs (6)&0.8009 & 0.001** & 0.001** & 0.0219*& 0.001**& 0.7343& 0.001** &0.001**& 0.001**\\
\hline \hline
\multicolumn{10}{l}{(1) Word2Vec features, (2)Mitchell's 25 features , (3) Glove features, (4) Meta-Embeddings features}\\
\multicolumn{10}{l}{(5) Randomly-generated features, (6) FastText features, **p$<$0.01, *p$<$0.05}\\
\end{tabular}
\end{table*}

\begin{figure*}[htbp]
\minipage{0.333\textwidth}
  \includegraphics[width=\linewidth]{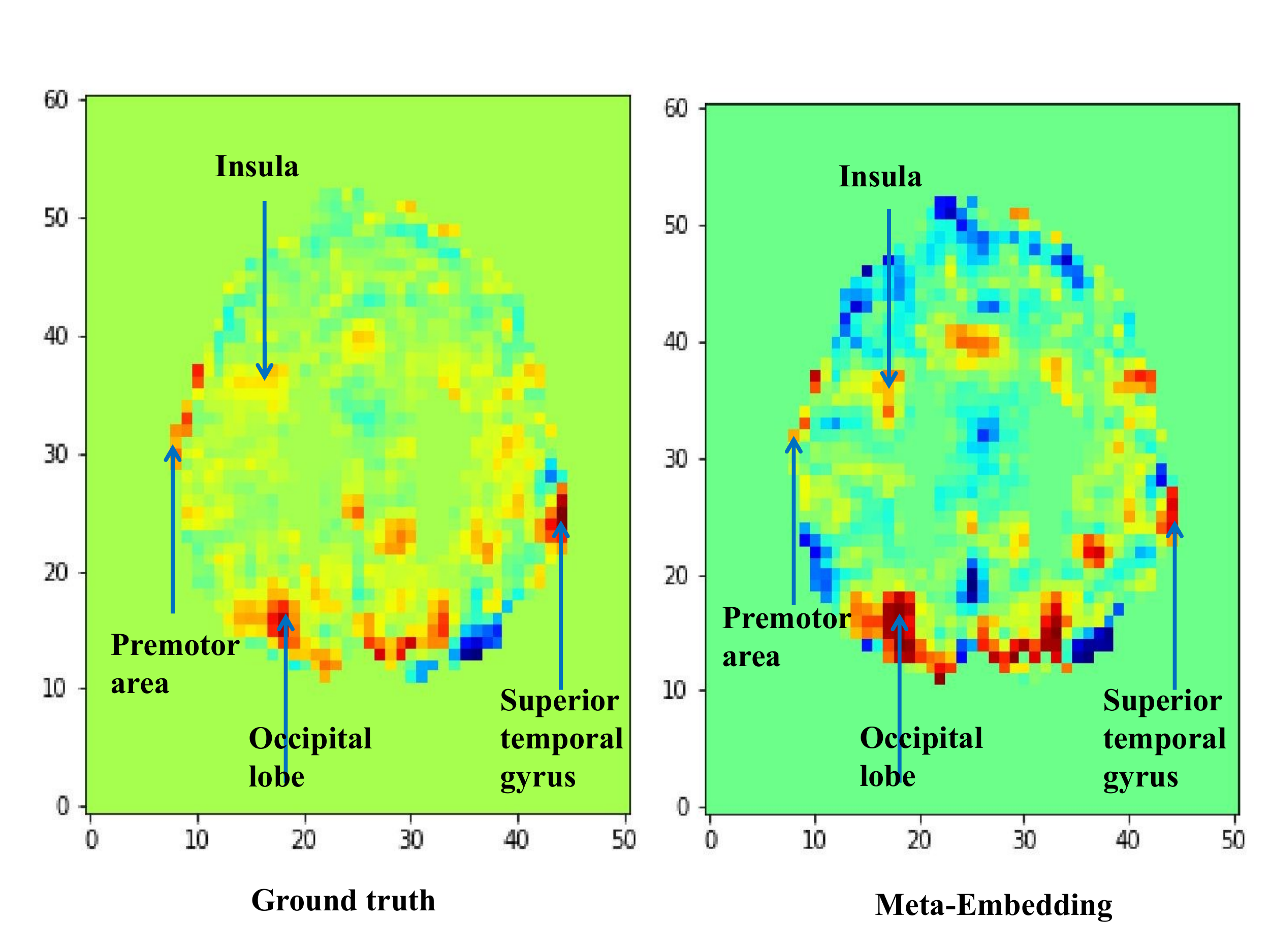}
  \vspace{-0.5cm}
\endminipage\hfill
\minipage{0.333\textwidth}
  \includegraphics[width=\linewidth]{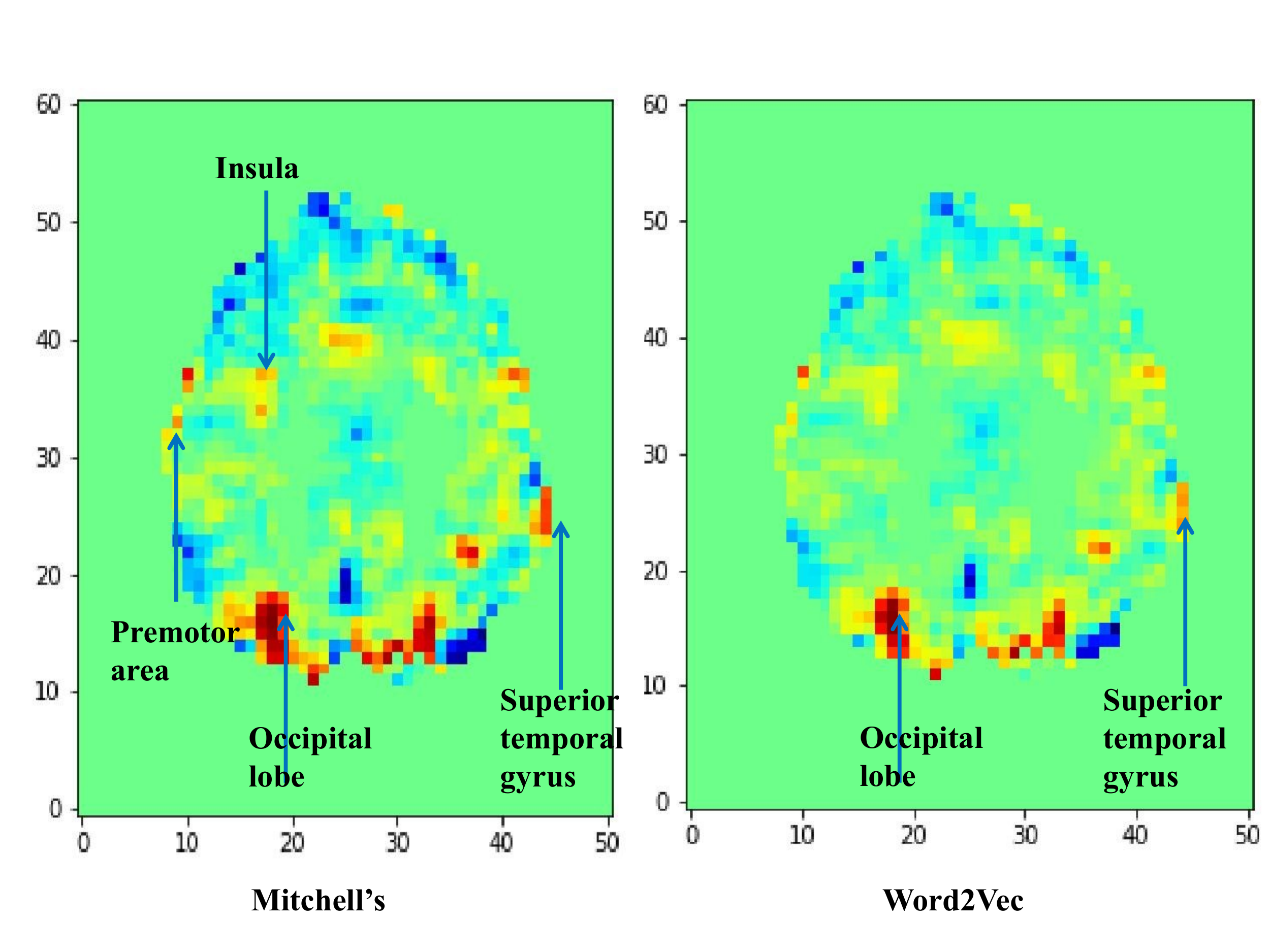}
  \vspace{-0.5cm}
\endminipage\hfill
\minipage{0.333\textwidth}
  \includegraphics[width=\linewidth]{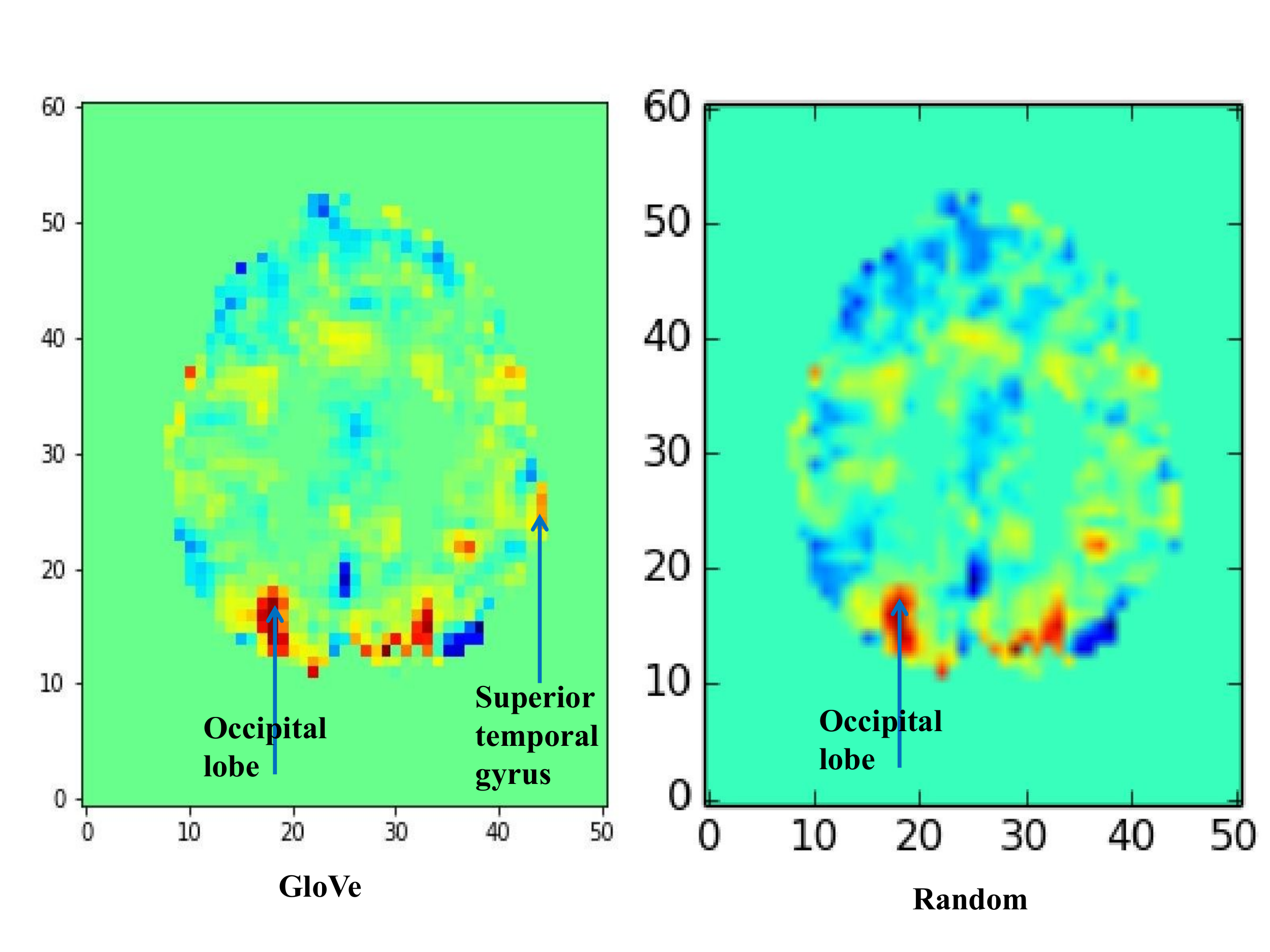}
  \vspace{-0.5cm}
\endminipage\hfill
\caption{Predicted fMRI image for the word ``Bicycle". One representative horizontal slice (taken at $z = 19$) for each method is displayed. From left to right: Ground truth, Meta-Embeddings, Mitchell's, Word2Vec, GloVe and Randomly generated features used to learn different decoding models.}
\label{fig:awesome_image4}
\end{figure*}

We use the rescaled mean squared error ($R^2$) as a metric to measure the error between predicted and target fMRI brain images. Kruskal-Wallis rank test~\cite{kachigan:statistical} was used for comparing mean ranks across the six methods in nine subjects. The one-way ANOVA test confirmed that there was a statistically significant difference between Meta-Embeddings, FastText, and Randomly generated vectors. Table~\ref{results} shows mean ranks of nine subjects when using six methods. From Table~\ref{results}, we can observe that Meta-Embeddings, GloVe, and Mitchell's features are not statistically significantly different from one another in all the nine subjects. This leads us to conclude that all these methods have similar performance. Word2Vec approach is statistically significantly different as compared to Mitchell's approach only in the case of subject-1 (see Table~\ref{results1}). The \emph{post-hoc} Scheffe's test~\cite{scheffe:method} results in Table~\ref{results1} show that $R^2$ values of Meta-Embeddings, GloVe, Mitchell's and Word2Vec differ significantly from those of the FastText vectors at $p$=0.001 and Random vectors at $p$=0.001. No significant differences were observed between mean ranks of the Meta-Embeddings, GloVe, Word2Vec and Mitchell's 25 features.

\subsection{Mapping Semantics onto the Brain}
To evaluate our computation model, we examine the fMRI signatures for the features used in six methods shown in Figure~\ref{fig:awesome_image4} for subject-2. These input features represent the model’s learned decomposition of neural representations into their component semantic features and depict substantial activities in different regions of the brain. From Figure~\ref{fig:awesome_image4}, we observe that predicted activations in multiple cortical regions using Meta-Embeddings approach seems similar to the state-of-the-art Mitchell's method. Some of the semantic features such as ``riding", ``see", ``say" and ``fear" associated with the word ``Bicycle" used in Mitchell's method lead to activations in the corresponding brain regions such as the ``Premotor Area", ``Occipital lobe / visual cortex", ``Superior temporal gyrus / auditory cortex" and ``Insula". In Meta-Embeddings, features like ``riding", ``spoke", ``surly"  associated with the word ``Bicycle" predicted similar activations as that of the Mitchell's method. However, the models using embedding methods such as Word2Vec and GloVe predicted activations only in the ``Occipital lobe / visual cortex" and ``Superior temporal gyrus / auditory cortex". Whereas the model using Randomly generated features failed to predict activations in the corresponding brain regions. 

\subsection{Statistical Analysis across subjects}
\begin{figure}[htbp]
\begin{center}
\includegraphics[width=0.8\textwidth]{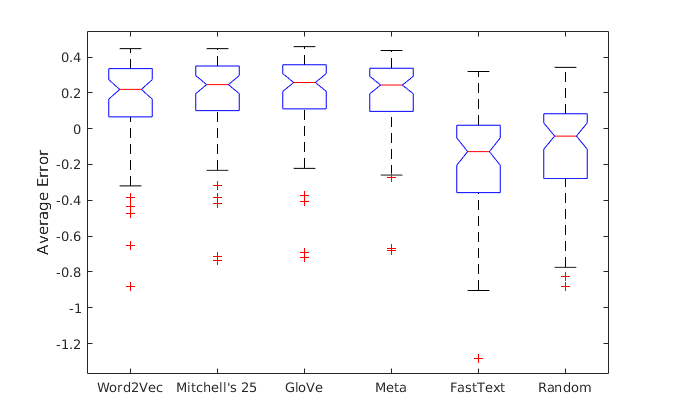}
\caption{ \textbf{Box-plot for average error of all subjects for six methods. Horizontal lines represent median ranks, notches represent 95\% confidence interval. Median rank of Random and FastText methods are significantly less than that of other four methods ($p$=0.0010053).}}
\label{box_plot}
\end{center}
\end{figure}

\vspace{-0.02cm}
Kruskal-Wallis rank test was used for comparing median
ranks across Word2Vec, GloVe, Meta-Embeddings, Mitchell's 25, Random and FastText methods performed across all subjects. The one-way ANOVA test confirmed that there was statistically significant
difference between average error of predicted fMRI image when using Word2Vec, GloVe, Meta-Embeddings, Mitchell's 25, Random and FastText methods ($p$=0.001) with a median rank of
0.2190 for Word2Vec, 0.2477 for GloVe, 0.2434 for Meta-Embeddings, 0.2459 for Mitchell's, -0.1281 for Random and -0.04139 for FastText. A
\emph{post hoc} Scheffe's test showed that average error of predicted fMRI image for Random and FastText methods differed significantly from those of the other four methods: Word2Vec, GloVe, Meta-Embeddings and Mitchell's 25 at $p$=0.0010053. No significant differences were observed
between median ranks of the other four word embedding methods. 
From Figure~\ref{box_plot}, we can observe that the average error for the models using embeddings Word2Vec, GloVe, Meta-Embeddings and features from Mitchell's 25 is similar and is significantly different from the average errors of the models using FastText and Random features.

\vspace{-0.2cm}
\vspace{-0.5cm}
\section{Conclusion}
This study employs the existing popular word embeddings such as Word2Vec, GloVe, Meta-Embeddings and FastText to scrutinize the semantic representations in brain activity as measured by fMRI. One of the main observations from our study was that while Mitchell's hand-crafted features were designed to cover multimodal activity of the brain covering several brain regions, the corpus-based word embedding models are based on word co-occurrence based statistics and thus lack the multi-modal context embedded in Mitchell's feature vector. Such general word embedding encoding schemes tend to give a strong within-category coverage for the input words and try to project this across different brain regions through associative mapping learned in the 3-layer neural network. Thus the current study can be considered a feasibility study of using generic word embedding schemes for brain decoding rather than painstakingly assembling hand-crafted features. Experimental results reveal that the $R^2$ error between Mitchell's approach and the other schemes such as Word2Vec, GloVe and Meta-Embeddings is small and the statistical significance of the results also points out that both the approaches are similar in their final outcome. In future, we would like to generate word embeddings from corpora from different (multi-modal) genres so that such feature vectors will also have an opportunity to learn mapping to multi-modal sensory and association regions of the brain. This might give us more insights into the mapping process of word-embedding representations to brain response and eventually improve the decoding accuracy of brain activation with such predictive solutions. The source code is publicly available at~\url{https://tinyurl.com/ya2v47bm} so that researchers and developers can work on this exciting problem collectively.

\bibliographystyle{splncs}
\bibliography{splncs}

\end{document}